\documentclass{aa}

\usepackage[varg]{txfonts}
\usepackage{subfig}
\usepackage{graphicx,natbib,bm,gensymb}
\usepackage[colorlinks,citecolor=blue,linkcolor=blue]{hyperref}

\graphicspath{{./}{figures/}}
\bibpunct[, ]{(}{)}{;}{a}{,}{,}

\newcommand{\revised}[1]{{#1}}

\begin{document}

\title{Global axisymmetric simulations of photoevaporation and magnetically driven protoplanetary disk winds}

\author{P. J. Rodenkirch\inst{1} \inst{2} \and
H. Klahr \inst{2} \and
C. Fendt \inst{2} \and
C. P. Dullemond \inst{1}}
\institute{Institute for Theoretical Astrophysics, Zentrum f\"ur Astronomie, Heidelberg University, Albert Ueberle Str. 2, 69120 Heidelberg, Germany \and
Max-Planck-Institut f\"ur Astronomie, K\"onigstuhl 17, 69117 Heidelberg, Germany}

\date{\today}

\abstract{Photoevaporation and magnetically driven winds are two independent 
mechanisms that remove mass from protoplanetary disks. In addition to accretion, 
the effect of these two principles acting concurrently could be significant, and 
the transition between them has not yet been extensively studied and quantified.}
{In order to contribute to the understanding of disk winds, we present the 
phenomena emerging in the framework of two-dimensional axisymmetric, nonideal 
magnetohydrodynamic simulations including extreme-ultraviolet (EUV) and X-ray driven photoevaporation.
  Of particular interest are the examination of the transition region between 
photoevaporation and magnetically driven wind, the possibility of emerging 
magnetocentrifugal wind effects, and the morphology of the wind itself, which 
depends on the strength of the magnetic field. }
{We used the PLUTO code in a two-dimensional axisymmetric configuration with additional 
treatment of EUV and X-ray heating and dynamic ohmic diffusion based on a 
semi-analytical chemical model. }
{We determine that the transition between the two outflow types occurs for values of the initial plasma beta $\beta \geq 10^7$, 
while magnetically driven winds generally outperform photoevaporation for 
stronger fields. 
In our simulations we observe irregular and asymmetric outflows for stronger 
magnetic fields. In the weak-field regime, the photoevaporation rates are 
slightly lowered by perturbations of the gas density in the inner regions of the 
disk. Overall, our results predict a wind with a lever arm smaller than 1.5, consistent with a hot magnetothermal wind. Stronger accretion flows are present for values of  $\beta < 10^7$.}{}


\titlerunning{Photoevaporation and MHD winds}

\maketitle

\section{Introduction}
\label{sec:intro}

Protoplanetary disks have an observed lifetime that ranges 
from 2 to 6 Myr \citep{Haisch2001, Mamajek2004, Ribas2015}. 
The physical processes that constrain and shape the evolution 
of these disks are highly debated. \\
One possible explanation for the limited lifetime is the phenomenon of 
accretion. 
In the study of accretion disks around black holes, \cite{Shakura1973} invoked 
the scenario that an $\alpha$ viscosity based on underlying turbulent effects 
drives the accretion flow. The concept can be applied to circumstellar disks, 
where several instabilities have been postulated over the last 
decades. \\
When the disk mass reaches a significant fraction of the stellar mass, gravitational instability (GI) \cite{Toomre1964} can operate, leading to 
spirals, fragmentation, and accretion flows.
 The nonlinear evolution of GI leads to a gravoturbulent state, depending 
on the balance of shock heating, compression heating, and cooling 
\citep{Durisen2007}, and it can be incorporated into $\alpha$ disk models 
\citep{Lin1987}. The critical cooling timescale for self-gravitating turbulence 
has been studied numerically by \cite{Lodato2004,Lodato2005} and 
\cite{Paardekooper2011}, for example. \\
A prominent mechanism that can explain turbulence-driven accretion is 
magnetorotational instability (MRI) \citep{Balbus1991}, which involves a well-ionized 
medium and a weak magnetic field embedded in a differentially rotating disk. 
Numerous studies on the convergence and influence of 
non-ideal magnetohydrodynamic (MHD) effects in the framework of local shearing 
box simulations have been published, for instance,
\cite{Fromang2007, Bai2011} and \cite{Hirose2011}. Global 
simulations including ohmic diffusion were carried out, for example, by 
\cite{Dzyurkevich2010} and \cite{Flock2011}. \\ 
Because the deeper layers of the disk are weakly ionized \citep{Igea1999}, ohmic 
diffusion dominates in the mid-plane, inhibiting MRI activity and thus creating 
a "dead zone" \citep{Gammie1996}. Its structure has been studied in global 
simulations by \cite{Dzyurkevich2013}.
In addition to MRI, hydrodynamic instabilities can potentially operate in the 
dead zone. Examples are vertical shear instability (VSI) \citep{Nelson2013}, 
Rossby-wave instability (RWI) \citep{Lovelace1999}, global baroclinic 
instability \citep{Klahr2003}, and convective overstability 
\citep{Klahr2014}.\\ 
As a different concept, a disk threaded by a large scale, an open magnetic field 
can develop magnetically (or magnetocentrifugally) driven winds that lead to 
angular momentum transfer and consequently cause accretion flows. A 
semi-analytical description of this concept has first been presented in the 
seminal paper of \cite{Blandford1982}. In order to allow for stationary solutions,
turbulent diffusivity has been assumed in the mid-plane of the disk 
\citep{Wardle1993, Ferreira1993, Ferreira1995}. These winds present a viable 
alternative solution to the $\alpha$ disk model.\\
\cite{Casse2002} presented the first global simulations in which a jet was launched from a 
resistive disk.
Similar simulations on a longer timescale have been reported by \cite{Zanni2007} 
with $\beta = 1$ in 
the mid-plane and extended ranges of viscosity and resistivity. 
\cite{Tzeferacos2009} observed unsteady winds up to $\beta \approx 
500$. \cite{Sheikhnezami2012} studied the effect of the magnitude and 
distribution of the magnetic diffusivity on the mass loading of the jet, and 
\cite{Fendt2013} focused on the symmetry of bipolar outflows. 
\\
\cite{Stepanovs2014} and \cite{Stepanovs2016} 
found the jet characteristics to be correlated to the disk magnetization and 
identified a transition between jets driven by magnetocentrifugal forces and magnetic 
pressure at $\beta \approx 100$. Furthermore, the disk magnetization can change substantially throughout the dynamical evolution of the outflow and the disk.\\
\cite{Gressel2015b} simulated global magnetically driven disk winds for 
weaker fields around $\beta = 10^5$ including ohmic and ambipolar diffusion. \cite{Bethune2017} included ohmic diffusion, ambipolar diffusion, and the 
Hall effect in magnetothermal wind simulations that included 
heating of the upper layers of the disk. The authors found asymmetric winds in some cases. 
\cite{Bai2017b} conducted a study that also included all three nonideal MHD effects with a comprehensive microphysical treatment. \\
In the framework of local shearing box simulations, \cite{Suzuki2009} found that a 
wind was launched from MRI turbulent layers at about two disk scale heights 
where the plasma beta $\beta = 8\pi P / B^2$ (the ratio of thermal to magnetic 
pressure) reaches unity, assuming $\beta$ to be $10^6$ in the mid-plane. A 
flow like this may create an inner hole in the disk \citep{Suzuki2010}. \\
In local simulations of \cite{Bai2013a}, the wind 
mass-loss rate was observed to be proportional to $1 / \beta$  up to a mid plane $\beta = 100$
.  
\cite{Bai2013b} stated that the MRI is suppressed at 1 au and that for $\beta = 
10^5$ a wind is launched at four disk scale heights. However, it was pointed out 
that the outflow rate may be sensitive to the applied resolution in local simulations \citep{Fromang2013}. \\
As an alternative to magnetically driven winds, outflows can be solely thermally 
driven through ionizing radiation from the central star or external sources. 
Possible types of radiation causing these photoevaporative winds are extreme-ultraviolet (EUV) 
radiation \citep{Hollenbach1994},  far-ultraviolet (FUV) radiation \citep{Adams2004, Gorti2004, 
Gorti2009} and X-ray radiation \citep{Ercolano2009, Owen2010, Owen2012, 
Picogna2019}. Pphotoevaporation does not drive accretion flows, therefore these winds 
merely act as a mass sink for the underlying disk.\\
The interplay between the two mechanisms of magnetically driven winds and 
photoevaporation has not been subject of detailed study in the past.
Only recently, a combination of these concepts has been presented by 
\cite{Wang2019}, involving global 2.5D magnetohydrodynamic simulations. In their 
model the simplified radiative transfer is evolved during each hydro step.\\
In 
this paper we use the recent photoevaporation model by \cite{Picogna2019} 
with a precomputed temperature prescription originating from detailed radiative 
transfer and photoionization calculations including EUV and X-ray radiation. We 
extend the fiducial model by applying a large-scale magnetic field. Thereby, we can  study the 
wind rates depending on the magnetic field strength in order to identify the 
transition region of magnetically driven winds and photoevaporation.\\
The paper is organized as follows: In section \ref{sec:theory} the general 
concept and theory of the two disk-wind mechanisms are outlined. Section 
\ref{sec:model} introduces the numerical model we used in our simulations, that is, the 
EUV and X-ray heating, ionization model, and disk properties. In section 
\ref{sec:results} we present the main results of our study, including wind flow, transition region, magnetic field evolution, and accretion flows. In sections 
\ref{sec:discussion} and \ref{sec:conclusion} these results are discussed and 
summarized.

\section{Theory}
\label{sec:theory}

\subsection{Magnetically driven outflows}
In general, magnetically driven outflows involve mass loss in the form of winds 
that are caused by a sufficiently ionized rotating disk threaded by a magnetic field. The wind-launching mechanism depends on the magnetic field strength, and the 
launching mechanism can be divided into two regimes with either strong or weak magnetic 
fields. These two regimes are magnetocentrifugal winds with strong 
magnetic fields and with winds that are solely driven by the magnetic pressure gradient if the 
magnetic field is weak. In between these extremes, a smooth 
transition exists at intermediate field strengths \citep{Bai2016}. \\
In the strong-field limit, the magnetic field lines rotate approximately rigidly up to the Alfv\'en surface, and the wind is magnetocentrifugally 
accelerated \citep{Blandford1982, Pudritz1983}. The condition for this wind to 
occur is that the magnetic field is inclined by less than $60 \degree$ 
to the disk. In this picture, the accelerated gas can be pictured like a string of pearls (the "string" being magnetic field lines). The inclination 
condition can be relaxed to $70 \degree$ when a hot wind is assumed \citep{Pelletier1992}. \\
Because we assume corotating field lines (i.e., the angular frequency along the field lines is approximately equal to the Keplerian angular frequency at 
the foot point of the wind), the gas parcels carry away angular momentum from 
the foot point at the disk. This results in an inflow of gas. \\
Accretion and wind-loss rates can be connected in the framework of magnetocentrifugally driven winds. 
The wind extracts a specific angular momentum of $j_w = \Omega (r_A - r_0)^2$ 
when it corotates up to the Alfv\'en radius $r_A$ starting from the foot point $r_0$.
The relation between the (cumulative) wind mass-loss rate $\dot{M}_\mathrm{w}$ and the accretion rate $\dot{M}_\mathrm{acc}$ can be derived as \citep{Ferreira1995, Bai2015b}
\begin{equation} \label{eq:lever}
\frac{\mathrm{d}\dot{M}_\mathrm{w}}{\mathrm{dln}\,r} = \frac{\dot{M}_\mathrm{acc}}{2} \frac{1}{\lambda^2 - 1}
,\end{equation}
with the so-called \textit{\textup{magnetic lever arm}} $\lambda$. \revised{The condition $\lambda \geq 1.5$ holds for cold MHD winds, and a smaller lever arm is possible when a warm or hot magnetothermal wind is assumed \citep{Bai2016}.} \\
Throughout this paper the cylindrical radius is denoted by $r$ and the vertical cylindrical coordinate is represented by $z$. The spherical radius $R$ is thus given by
\begin{equation} \label{eq:spherical_radius}
R^2 = r^2 + z^2.
\end{equation}
In the strong magnetic field limit the outflow can collimate to a jet 
\citep{Casse2002}.
With decreasing magnetic field strengths, the wind flows in a wider angle and 
becomes increasingly driven by the magnetic pressure gradient, which is in contrast to the 
centrifugal acceleration with a strong magnetic field. In the context of this 
"magnetic tower wind" \citep{Lynden-Bell1996, Lynden-Bell2003}, the toroidal 
magnetic field is dominant and the Alfv\'en surface lies close to the wind-launching front. Consequently, the angular momentum transport is small and the 
accretion rate decreases in the disk, which leaves the wind as the dominant cause for mass 
loss. Moreover, in the transition from strong to weak magnetic fields, the wind 
topology changes from a jet-like structure to a more unsteady, episodic wind 
\citep{Sheikhnezami2012}.

\subsection{Photoevaporation and X-Ray heating} 
\label{sec:intro_photoevaporation}

The fundamental idea of photoevaporation is that high-energy radiation ionizes 
parts of the circumstellar disk and thereby heats the topmost disk layers. These hot layers subsequently expand and produce a pressure-driven 
transonic wind, similar to a Parker wind.
The mechanism can be divided into external photoevaporation, where the ionizing 
radiation originates from external source (e.g., O stars) and internal 
photoevaporation, where the central star provides the driving radiation. 
We consider only the latter here. \\
The energy of the incoming radiation can be categorized into three regimes: 
FUV  (6 - 13.6eV), EUV (13.6 - 100eV), and X-ray (0.1 - 100keV) 
radiation.
Typical mass-loss rates expected from photoevaporation are about $3 
\cdot 10^{-8} \, \text{M}_{\odot} \text{yr}^{-1}$ \citep{Gorti2009}. \\
In models of EUV-driven photoevaporation, the photon flux ionizes hydrogen atoms 
in the upper atmosphere layers of the protoplanetary disk. Thereby, the created 
diffuse recombination radiation heats up the outer regions of the disk 
\citep{Hollenbach1994,Clarke2001,Font2004}, leading to mass-loss rates of 
$\approx 4 \cdot 10^{-10} \, \text{M}_{\odot} \text{yr}^{-1}$.
 A characteristic length scale of these systems is the so-called 
\textit{\textup{gravitational radius,}}
\begin{equation}
R_g = \frac{G M_*}{c_s^2}
,\end{equation}
where $c_s$ is the isothermal sound speed, $G$ is the gravitational constant, and 
$M_*$ is the stellar mass.
With ionization temperatures of $\approx 10^4$K
and assuming a solar mass star, $R_g$ is $\approx 9\,\mathrm{au}$.  
However, more detailed studies \citep{Adams2004,Font2004} have shown that the 
sharp division, induced by the gravitational radius, is in fact a more diffuse 
boundary, situated at roughly 0.1 - 0.2 $R_g$ (\textup{critical radius}) 
\citep{Dullemond2007}. 
 In the long-term evolution of photoevaporation in combination with viscous 
evolution, a two-timescale behavior emerges. This creates a gap at $R_g$ when 
the local accretion rate is comparable to the wind mass-loss rate. 
Then, the inner disk quickly accretes onto the star within about $10^5$ yr 
\citep{Clarke2001}. \\
With an inner hole, direct illumination of the outer part leads to thermal sweeping. This clears the remaining disk in an inside-out 
fashion \citep{Owen2012, Haworth2016}. \\
X-ray photons are able to penetrate deeper into the disk atmosphere than EUV 
radiation (column densities of $\approx 10^{22} \, \mathrm{cm}^{-2}$ 
\citep{Ercolano2009,Owen2010}). 
\cite{Owen2010}
argued that X-ray driven photoevaporation leads to significantly 
higher mass-loss rates than the EUV case. Models of 
\cite{Owen2010,Owen2012} indicate a mass-loss rate of $1.4 \cdot 10^{-8} \, 
\mathrm{M}_{\odot} \text{yr}^{-1}$ for solar-like stars, involving the concept 
of an ionization parameter, 
\begin{equation}
\label{eqn:ionparam}
\xi = \frac{L_x}{n \cdot R^2}
,\end{equation}
 where $L_x$ denotes the luminosity of the central X-ray radiation source and 
$n$ the local number density of the gas. 
By computing a relation between $\xi$ and the local gas temperature, tabulated 
values can be interpolated locally during the hydrodynamical simulation because 
in-place calculations of the temperatures would be computationally prohibitive. 
The main assumption here is that we are in the optically thin regime and hence 
X-ray attenuation can be neglected up to the limiting column density.Defining a gravitational radius in this case is less useful because the gas 
temperatures depend on the density, and the heated layers cannot be assumed to be 
isothermal.\\
Far-ultraviolet radiation can heat the ambient medium by photoelectric heating of grains as well 
as by vibrational excitation and collisional relaxation of $\mathrm{H}_2$ molecules 
\citep{Hollenbach1999}. \cite{Gorti2009}
found an FUV-dominated mass-loss rate of $\approx 3 \cdot 10^{-8}\mathrm{M}_\odot 
\mathrm{yr}^{-1}$ at a distance range of 100 au to 200 au, which is comparable to 
X-ray photoevaporation. However, the temperatures in FUV-heated regions are 
sensitive to the uncertain abundance of polycyclic aromatic hydrocarbons (PAHs) 
\citep{Geers2009}. Because of these drawbacks and the chemical complexity, we 
restrict our model to EUV and X-ray dominated heating.

\section{Numerical model} 
\label{sec:model}

All simulations of this paper are two-dimensional axisymmetric, including the 
toroidal velocity and magnetic field component (i.e., 2.5 dimensional). They were carried out with the PLUTO code (version 4.3) 
\citep{Mignone2007,Mignone2012}, which allows solving the MHD equations on a 
spherical grid, including ohmic resistivity. Throughout this section, we present the basis 
of the numerical model, that is, the X-ray and EUV heating approach, disk 
structure, magnetic field, and dynamic treatment of ohmic diffusion. 

\subsection{MHD equations}
Relevant equations are the conservation of mass,
\begin{equation}
\frac{\partial \rho}{\partial t} + \nabla \cdot (\rho \bm{v})
,\end{equation}
the conservation of momentum,
\begin{equation}
\frac{\partial (\rho \bm{v})}{\partial t} + \nabla \cdot \left(\rho \bm{v} 
\bm{v}^T - \frac{\bm{B}\bm{B}^T}{4\pi} \right) + \nabla \left( P + 
\frac{B^2}{8\pi} \right) + \rho \, \nabla \Phi = 0
,\end{equation}
the energy equation,
\begin{equation} \label{eq:energy}
\frac{ \partial e}{\partial t} + \nabla \cdot \left[ \bm{v} \left(e + P + 
\frac{B^2}{8 \pi}\right) - \frac{1}{4 \pi} (\bm{v} \cdot \bm{B}) \bm{B} + \left( \eta \cdot \bm{J} \right) \times \bm{B} \right] 
= - \rho (\nabla \Phi) \cdot \bm{v}
,\end{equation}
and the induction equation,
\begin{equation}
\frac{\partial \bm{B}}{\partial t} - \nabla \times \left(\bm{v} \times \bm{B} - 
\eta \cdot \bm{J} \right) = 0
,\end{equation}
where $P$ is the thermal gas pressure, $\rho$ the gas density, $\bm{B}$ the 
magnetic flux density vector, $\bm{J}$ the electric current density vector, 
$\eta$ the ohmic diffusion coefficient, $\bm{v}$ the gas velocity vector, and 
$\Phi = -\frac{GM_*}{R}$ the gravitational potential. \\
To update the cells, the \textit{HLLD} Riemann solver \citep{Miyoshi2005} was 
used. To ensure the condition $\nabla \cdot \bm{B} = 0,$ the divergence 
cleaning method \citep{Dedner2002} was applied.
All simulations were evolved with an adiabatic equation of state, using an isentropic exponent of $\kappa = 5/3$.

\subsection{Diagnostics}

The relative strength of the magnetic field can be expressed with the 
dimensionless plasma parameter $\beta$:
\begin{equation}
\beta = \frac{8 \pi P}{B^2}
,\end{equation}
which is the ratio of thermal pressure over magnetic pressure. \\
In order to evaluate the non-ideal MHD effects (ohmic diffusion) on 
possible MRI activity, the notion of an ohmic Elsasser number $\Lambda_\Omega$ 
\citep{Turner2007}  is useful:
\begin{equation}
\Lambda_\Omega = \frac{v_A^2}{\eta \, \Omega_K}
.\end{equation}
Here $v_A$ is the Alfv\'en velocity and $\eta$ the ohmic diffusion coefficient. 
Generally, MRI is considered to be suppressed for $\Lambda_\Omega \lesssim 
1$. \\

The integrated mass-loss rate caused by the wind is computed by the 
expression
\begin{equation}
\dot{M}_{\mathrm{w}} =  2 \pi \int \rho v_R \, r^2 \mathrm{sin}(\theta) 
\mathrm{d} \theta
,\end{equation}
and the disk region is excluded from the range in $\theta$. Here, $v_R$ 
denotes the radial velocity component of the flow.

\subsection{EUV and X-ray heating} 
\label{sec_photoevaporation}

Following the approach of \cite{Owen2010,Owen2012,Owen2013}, X-ray heating was 
provided by using a fitted ionization parameter - temperature relation $T = 
f(\xi)$, taken from \cite{Picogna2019}.
\begin{figure}
\includegraphics[width=\linewidth]{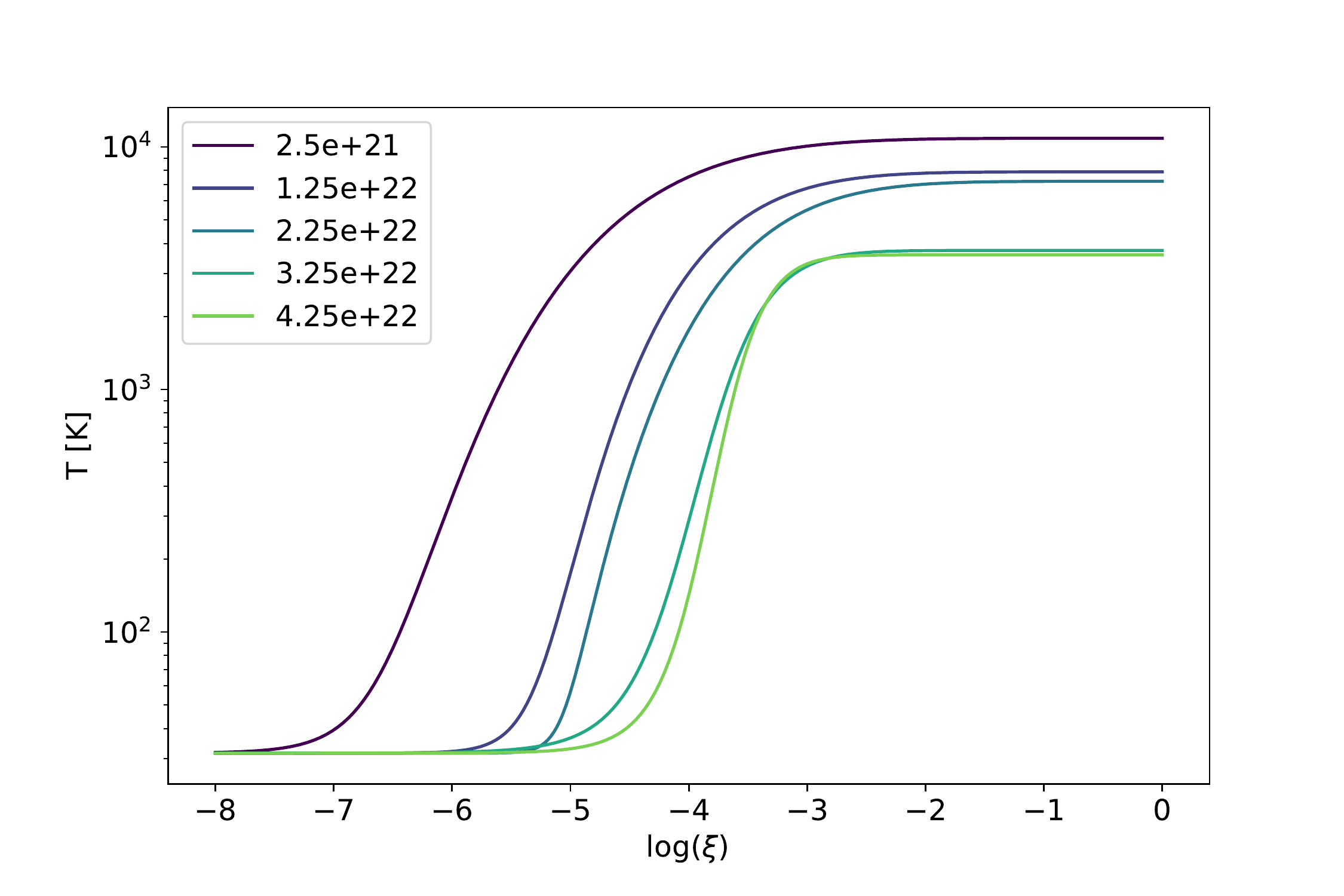}
\caption{Plot of the ionization parameter - temperature relation based on the 
data in \cite{Picogna2019}
 \label{fig:heating}}
\end{figure}
 In addition to the previously studied models, the temperature curves depend on 
the local column density, as shown in Fig. \ref{fig:heating}.
Knowing the dependence of the temperature on the local variables, we traced rays 
from the inner simulation boundary radially outward up to column densities of $5 
\cdot 10^{22} \, \mathrm{cm}^{-2}$. Within this range, the temperatures were set 
corresponding to the locally computed ionization parameter using the fitted 
relation $f(\xi)$. With intermediate column densities, the temperature is 
linearly interpolated.  \\
The temperature adjustment was only invoked when the temperature computed by the 
ionization model was higher than the local gas temperature. No relaxation time was 
applied. The regions that were unaffected by the photoevaporation heating were treated according to the energy equation (eq. \ref{eq:energy}). The resulting pressure in the simulations 
depends on the mean molecular weight $\mu$. Similarly to the models of 
\cite{Owen2010}, we set it to $\mu = 1.37125,$ which is suitable for ionized gas, 
including atomic hydrogen. \\
The implementation, depending on the column density, provides a more accurate 
treatment of the regime between optically thin and thick gas, where X-ray 
attenuation becomes significant. The temperature 
parameterization was computed by 3D photoionization and radiative transfer 
calculations. Thus the diffuse, secondary radiation field was taken into account. Essentially, the same initial parameters for the luminosity and 
the photon energy can be used to assemble a consistent model regarding the 
ionization rate for the ohmic diffusion coefficients (cf. section 
\ref{sec_ionrate}).

\begin{figure*}[ht]
\centering
\subfloat[]{ \label{photoev_vfield}
        \includegraphics[width=0.5\textwidth]{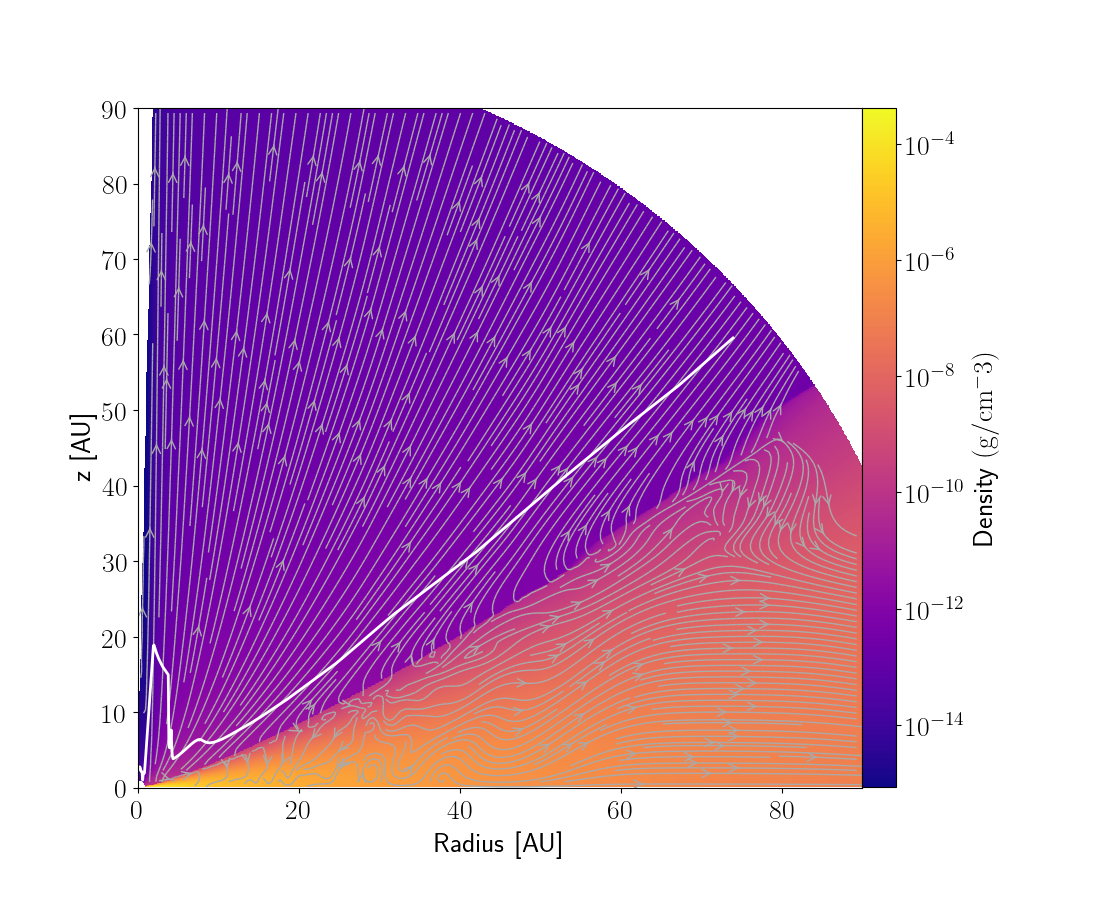}
}
\subfloat[]{ \label{photoev_massflux}
        \includegraphics[width=0.5\textwidth]{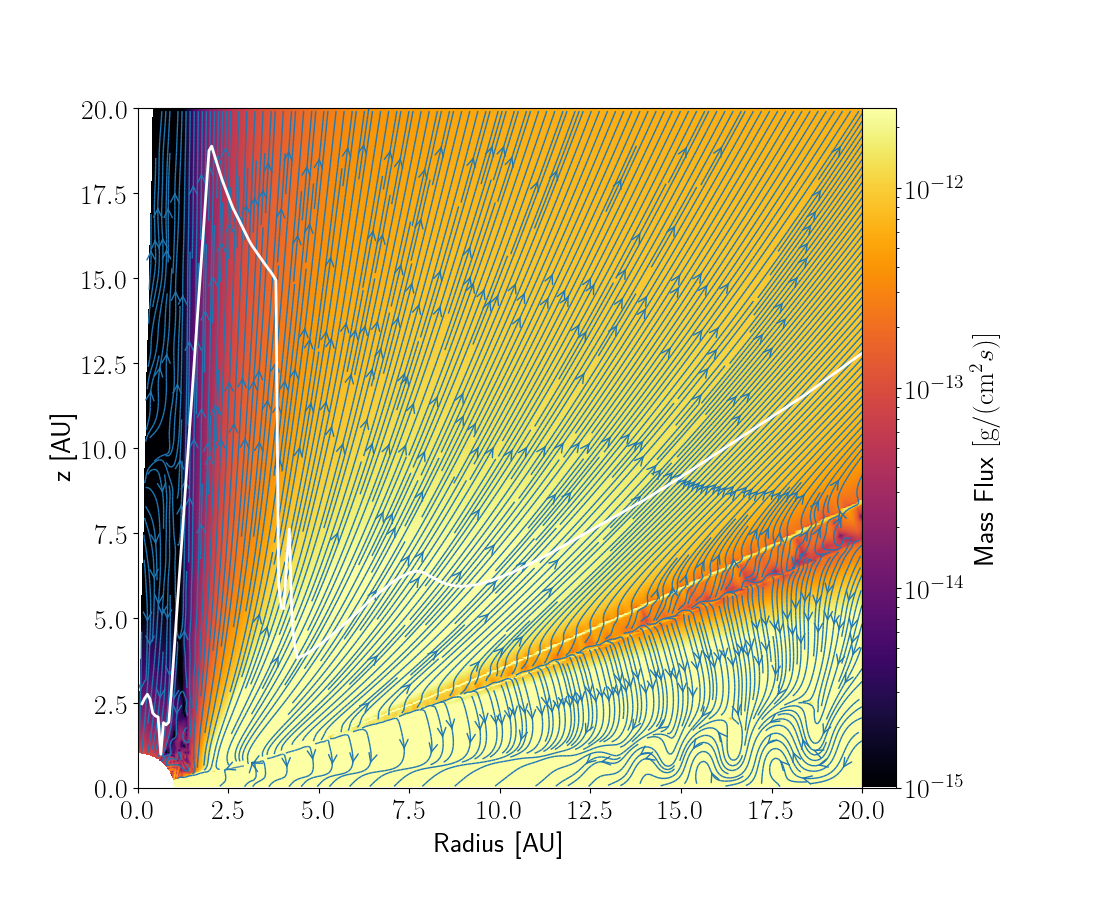}
}
\caption{Both panels are based on the simulation X-bn-h, time averaged from 555 
to 714 years. The white line represents the sonic surface in the wind flow, and 
the blue field lines correspond to the velocity stream lines. In figure 
\ref{photoev_massflux} the mass flux $\rho \boldmath{v}^2$ is shown. Within 
$\approx 3\,\mathrm{au,}$ a significantly lower mass flux is visible. The photoevaporation simulations are only carried out for one hemisphere.}
\label{fig:photoev_avg}
\end{figure*}

\subsection{Disk model}

\subsubsection{Initial density structure}

The initial density structure we used in the simulations can be formulated as
\begin{equation}
\Sigma(r) = 711 \frac{\mathrm{g}}{\mathrm{cm}^2} \left(\frac{r}{\mathrm{au}} 
\right)^{-1},
\end{equation}
where $\Sigma(r)$ is the vertically integrated column density. The scaling with 
the cylindrical radius $r$ corresponds to a column density gradient of $p = 1$. 
When we assume an initial disk mass of $\approx 0.05 M_\odot$ , the value of 
$\Sigma(r = 1 \, \mathrm{au})$ is equal to $\approx 711 \mathrm{g} / 
\mathrm{cm}^2$.  

For the initial thermal structure, the prescription from \cite{Hayashi1981} was 
used,
\begin{equation} \label{eq:temperature_midplane}
        T_{\mathrm{mid}}(r) = 280 K \left(\frac{r}{1 \, 
\mathrm{au}}\right)^{-\frac{1}{2}} 
\left(\frac{L_*}{L_{\odot}}\right)^{\frac{1}{4}}
,\end{equation}
which corresponds to a temperature gradient of $q = 1/2$ and causes the disk to 
be flared ($H / r$ not constant with radius). The dust temperature was assumed to 
be equal to the gas temperature in our model.\\
With the assumption of hydrostatic equilibrium and an isothermal, vertically 
stratified disk atmosphere, the density can be written as
\begin{equation} \label{eq:density_stratified}
\rho(r, z) = \rho_\mathrm{mid} \, \mathrm{exp} \left[ \frac{r^2}{H^2} \left( \frac{r}{R} - 
1  \right) \right]
,\end{equation}
where $H = c_s / \Omega_K$ is the pressure scale height and $\rho_\mathrm{mid}$ is the mid-plane density at $z = 0$. \\
For $z \ll r$ ,  Eq. \ref{eq:density_stratified} reduces to the 
conventional form $\rho_\mathrm{mid} \, \mathrm{exp}(-z^2/ (2H^2))$. However, Eq. 
\ref{eq:density_stratified} was used in our simulations because the disk has a 
significant vertical extent. 
The isothermal sound speed is given by $c_s = \sqrt{k_B T / \mu m_\mathrm{p}}$. We chose 
the mean molecular weight to be $\mu = 1.37125$, equal to the weight of the 
X-ray heated gas. \\
The density in the mid-plane is calculated with the column density by $\rho_\mathrm{mid} = 
\Sigma / \sqrt{2 \pi} H$. 
In a more general setting with surface density scaling as $\Sigma \propto r^p$ 
and temperature scaling as $T_{\mathrm{mid}} \propto r^q$ , we find that 
$\rho_\mathrm{mid}$ scales proportionally to $r^{-p + (3-q) / 2}$ (because $H \propto 
r^{(3-q)/2}$). \\
In protoplanetary disks, the toroidal velocity has to be lower than Keplerian 
because of the additional pressure force, which points outward. When we take 
vertical variations into account, the velocity in azimuthal direction can be defined as (using 
eq.(9) of \cite{Bai2017b})
\begin{equation}
v_\phi(r, R) = r \Omega_K \left[ 1 - (2p + (3-q)/2) \frac{H^2}{r^2} + q \frac{r 
- R}{R} \right]
.\end{equation}
By taking these vertical corrections into account, an initially more stable disk 
configuration can be achieved in the simulations. In order to verify the 
stability, the setup was tested without photoevaporation effects and magnetic 
fields.

\subsubsection{Coronal gas structure} \label{sec:corona}
The whole disk in the simulations is surrounded by a hydrostatically stable 
corona, consisting of gas whose density is much lower than that of the disk. Its 
purpose is to define a reasonable density floor value for numerical stability. 
When the mass of the sphere itself is neglected, the hydrostatic balance can be written as
\begin{equation}
\rho_c(R) = \rho_{c0}(r_i) \left( \frac{r_i}{R} \right)^{\frac{1}{\gamma-1}}
,\end{equation}
where $r_i$ is the radius at the inner boundary of the simulation domain and 
$\gamma$ is the adiabatic index ($\gamma = 5 / 3$ in the simulations). The density 
at the inner boundary $\rho_{c0}(r_i)$ is related to the disk density in the 
mid-plane and $r_i$ with the density contrast $\delta = \rho_{c0}(r_i) / 
\rho_\mathrm{mid}(r_i)$. The definition of the corona is the same as in \cite{Sheikhnezami2012}. The pressure can be conveniently scaled with the 
Keplerian velocity at the inner boundary,
\begin{equation}
P_c(R) = \rho_{c0}(r_i) \frac{\gamma - 1}{\gamma}\frac{GM_\odot}{r_i} 
\left(\frac{r_i}{R} \right)^{\frac{\gamma}{\gamma -1}}
.\end{equation} 
 In the simulations performed here, the upper limit of this density contrast was 
heavily constrained by the photoevaporation because the coronal gas should allow 
rays to traverse the whole simulation domain, without being absorbed completely 
or affected significantly. We therefore set the density contrast to $\delta = 
10^{-8}$.
The corona is mostly blown away by the emerging wind, but we find the solution 
to be numerically more stable with this prescription and its application as a 
density floor at every integration step. By updating the density to the floor 
value, the conservation of momentum is ensured.

\subsubsection{Magnetic field}

The initial configuration of the magnetic field was based on the publication of 
\cite{Zanni2007}. However, the flux function was slightly modified for PLUTO and 
our disk properties. A reasonable choice of the magnetic field strength in 
radial direction would be a power law, which results in a plasma beta that is constant 
in radius. Because the pressure in the chosen disk model scales with $r^{-11/4}$, 
the magnitude of the magnetic flux density should be proportional to 
$r^{-11/8}$. Then, the $\phi$-component of the vector potential is
\begin{equation} \label{eq:vector_potential}
A_3 = -\frac{8}{3} B_{z0} r_i \left( \frac{r}{r_i} \right)^{-\frac{3}{8}} \, 
\frac{m^{\frac{5}{4}}}{(m^2 + (\frac{z}{r})^2)^{\frac{5}{8}}}
,\end{equation}
with $B_{z0}$ the magnetic field strength in the vertical direction at the mid 
plane. The radial and toroidal magnetic field are zero in the mid-plane.
The parameter $m$ controls the bending of the magnetic field, where $m 
\rightarrow \infty$ would lead to a homogeneous vertical field. By applying 
$\nabla \times \bm{A}$, we obtained the magnetic field, where the first and 
second components would be zero in this case (no initial toroidal magnetic 
field). 

\subsubsection{Ionization rate} 
\label{sec_ionrate}

For the X-ray ionization rate $\zeta_\mathrm{X}$ we used a fitted relation by 
\cite{Igea1999} in which the direct radiation and the scattered secondary 
component are included,
\begin{flalign}
\zeta_X = & \, \zeta_{\mathrm{X}, \mathrm{sca}} \left[ \mathrm{exp} \left( - 
\frac{\Sigma_{\mathrm{top}}}{\Sigma_1} \right)^{0.65} +\mathrm{exp} \left( - 
\frac{\Sigma_{\mathrm{bot}}}{\Sigma_1} \right)^{0.65}   \right] \left( 
\frac{R}{\mathrm{au}} \right)^{-2} .\\
& + \zeta_{\mathrm{X}, \mathrm{rad}} \, \mathrm{exp} \left( 
\frac{\Sigma_\mathrm{rad}}{\Sigma_2} \right)^{0.4} \left( \frac{R}{\mathrm{au}} 
\right)^{-2} \notag
\end{flalign}
The coefficients for the scattered ionization rates and the column densities are 
$\zeta_{\mathrm{X}, \mathrm{sca}} = 2\cdot 10^{-14} \, \mathrm{s}^{-1}$ and $\Sigma_1 = 
7 \cdot 10^{23} \mathrm{cm}^{-2}$. The values for the direct, radial component 
are $\zeta_{\mathrm{X}, \mathrm{rad}} = 1.2 \cdot 10^{-10} \, \mathrm{s}^{-1}$ and 
$\Sigma_2 = 1.5 \cdot 10^{21} \mathrm{cm}^{2}$.
Cosmic rays are included by the relation stated in \cite{Umebayashi2009},
\begin{flalign}
\zeta_\mathrm{cr} = \, 5 \cdot 10^{-18} \mathrm{s}^{-1} & \mathrm{exp} \left( - 
\frac{\Sigma_{\mathrm{top}}}{\Sigma_\mathrm{cr}} \right) \left[ 1 + \left( 
\frac{\Sigma_\mathrm{top}}{\Sigma_\mathrm{cr}} \right)^{3/4} \right]^{-4/3}  \\
+ & \mathrm{exp} \left( - \frac{\Sigma_{\mathrm{bot}}}{\Sigma_\mathrm{cr}} 
\right) \left[ 1 + \left( \frac{\Sigma_\mathrm{bot}}{\Sigma_\mathrm{cr}} 
\right)^{3/4} \right]^{-4/3}, \notag %\right\rbrace 
\end{flalign}
where $\Sigma_\mathrm{cr} = 96 \, \mathrm{g} / \mathrm{cm}^2$. \\
To account for the ionization of short-lived radio nucleides in the disk, we added 
an ionization rate of $\zeta_\mathrm{nuc} = 7 \cdot 10^{-19} \, \mathrm{s}^{-1}$ 
caused by $^{26}\mathrm{Al}$ \citep{Umebayashi2009}.
The total ionization rate is then simply $\zeta = \zeta_\mathrm{X} + 
\zeta_\mathrm{cr} + \zeta_\mathrm{nuc}$.
In addition to the X-ray ionization, we also took the effect of FUV 
ionization into account using a simple prescription from eq. 17 in \cite{Bai2017b}, 
\begin{equation}
x_{e, \mathrm{FUV}}=2.0 \times 10^{-5} \exp \left[-\left(\frac{\Sigma_{r}(r, 
\theta)}{\Sigma_{\mathrm{FUV}}}\right)^{4}\right],
\end{equation}
which depends on the radial column density $\Sigma_{r}(r, \theta)$ and a 
critical column density $\Sigma_{\mathrm{FUV}} = 0.03 \, \mathrm{g} / 
\mathrm{cm}^2$. The wind rates are found to remain much the same when FUV ionization is included.\\

\subsubsection{Ohmic diffusion} \label{sec_ohmic_diffusion}

\begin{table*}[t]
\caption{Simulation parameters}
\centering
\begin{tabular}{lcccccr}
\hline
Label & $\beta_0$ & X-ray heating & $R_{\mathrm{in}}$ [au] & $R_{\mathrm{out}}$ [au] & Resolution & Simulation time\\
\hline
\hline
X-b5 & $10^5$ & $\checkmark$ & 1 & 60 & 400 x 620 & 6671 $\Omega^{-1}$\\
X-b6 & $10^6$ & $\checkmark$ & 1 & 60 & 400 x 620 & 10000 $\Omega^{-1}$\\
X-b6-mri & $10^6$ & $\checkmark$ & 1 & 60 & 200 x 200 & 10000 $\Omega^{-1}$\\
X-b7 & $10^7$ & $\checkmark$ & 1 & 60 & 400 x 620 & 10000 $\Omega^{-1}$\\
X-b8 & $10^8$ & $\checkmark$ & 1 & 60 & 400 x 620 & 10000 $\Omega^{-1}$\\
X-b9 & $10^9$ & $\checkmark$ & 1 & 60 & 400 x 620 & 10000 $\Omega^{-1}$\\
X-b10 & $10^{10}$ & $\checkmark$ & 1 & 60 & 400 x 620 & 10000 $\Omega^{-1}$\\
X-bn & \ldots & $\checkmark$ & 1 & 60 & 400 x 620 & 10000 $\Omega^{-1}$\\
X-bn-h & \ldots & $\checkmark$ & 1 & 100 & 400 x 580 & 70000 $\Omega^{-1}$\\
\hline
\end{tabular}
\tablefoot{All labels starting with X belong to simulations with X-ray heating 
enabled. The resolution is represented as the number of cells in radial and 
polar ($\theta$) direction, ranging from $\theta_b = 0.005$ to $\theta_e = \pi - 
0.005$ with the exception of run X-bn-h, where $\theta_e$ equals $\pi / 2$. The 
simulation time is measured in units of the orbital time scale $\Omega^{-1}$ at 
$1\,\mathrm{au}$. The time is converted into actual orbits by dividing the 
value by $2 \pi$.} \label{tab:simulations}
\end{table*}
The ohmic diffusion coefficient $\eta$ was computed at every grid cell and time 
step based on the semi-analytical model of \cite{Okuzumi2009}. The model is based on 
fractal aggregates and the network reduces to a root-finding problem of a single 
analytic expression. We solved eq. 34 in \cite{Okuzumi2009},
\begin{equation}
\frac{1}{1 + \Gamma} - \left[ \frac{s_\mathrm{i} u_\mathrm{i}}{s_\mathrm{e} 
u_\mathrm{e}} \mathrm{exp} \, \Gamma + \frac{\Gamma}{\Theta} \right],
\end{equation} 
where $\Theta$ is
\begin{equation}
\Theta = \frac{\zeta n_\mathrm{g} e^2}{s_\mathrm{i} u_\mathrm{i} \bar{\sigma} 
\bar{a} n_\mathrm{d}^2 k_\mathrm{B} T}
,\end{equation}
with the local gas density $n_\mathrm{g}$, the dust density $n_\mathrm{d}$, the 
ion and electron
%\LEt{please check what you mean with the slash here and below, "and", "or", or something else?}
 sticking coefficient $s_\mathrm{i}$, $s_\mathrm{e}$, the ion and 
electron thermal velocity $u_\mathrm{i}$, $u_\mathrm{e}$, the averaged grain 
cross section $\bar{\sigma,}$ and the grain size $\bar{a}$. The solution depends 
on the dimensionless variable $\Gamma = (- \langle Z \rangle e^2) / 
(k_\mathrm{B} T)$. When the solution is obtained, the electron density $n_e$ and 
thereby the ionization fraction $x_e = n_e / n_g$ can be calculated by 
\begin{equation}
n_e  = \frac{\zeta n_\mathrm{g}}{s_\mathrm{e} u_\mathrm{e} \bar{\sigma} 
n_\mathrm{d}} \mathrm{exp} \, \Gamma
.\end{equation}
We assumed grain aggregates with 400 monomers, resulting in a size of 2 $\mu 
\mathrm{m}$ and a dust-to-gas ratio of $f = 10^{-2}$. The sticking coefficients 
$u_\mathrm{i}$ and $u_\mathrm{e}$ were set to 1 and 0.3, respectively 
\citep{Okuzumi2009}. Furthermore, we assumed icy grains with a density of $1.4 
\mathrm{g} / \mathrm{cm}^2$.

With the ionization fraction $x_e$, the ohmic diffusion coefficient $\eta$ can 
be calculated following \cite{Blaes1994},
\begin{equation}
\eta = \frac{234}{x_e} \, T^{\frac{1}{2}} \text{cm}^2 \text{s}^{-1}
.\end{equation}
This expression is only valid when the conductivity of charged grains is 
negligible compared to the electron conductivity \citep{Wardle2007}. The 
diffusion coefficient $\eta$ may be less accurate deep within the disk toward the 
mid plane. However, we are mainly interested in the dynamics of the upper regions 
of the disk, and thus we used this approximate formulation. \\
In dimensionless code units the value of the ohmic diffusion coefficient at the inner boundary in the mid-plane initially reaches $\eta \approx 8 \cdot 10^{-3}$.

\subsection{Boundary conditions}
In the inner radial boundary all primitive variables at the boundary were copied 
into the ghost cells, except for the radial velocity component $v_R$ , for which the 
condition $v_R^{\mathrm{ghost}} = \mathrm{min} \left( v_R^{\mathrm{IBEG}}, 0 
\right)$ was applied, where $v_R^{\mathrm{ghost}}$ are the radial 
velocities in the ghost domain and $v_R^{\mathrm{IBEG}}$ the radial velocities 
on the active hydro mesh at the inner radial boundary. We thus avoided 
artificial inflow of gas from the inner 
radial boundary, which could lead to spurious effects in the domain. 
Similar to 
the inner boundary, the outer radial boundary follows the outflow prescription 
defined above,
thus avoiding infall of gas from the outer boundary region.
In order to avoid artificial collimation at the radially outer boundary, the 
value of $B_\phi$ was linearly extrapolated $\propto 1/R$ into the ghost cells. 
Both boundaries in $\theta$ direction were set to axisymmetric boundary 
conditions, that is, normal and azimuthal velocity and magnetic components flip sign.

\subsection{Simulation parameters and normalization} 
\label{sec:pluto_simulations}
Internal code units in PLUTO were scaled by the following normalization factors: 
\begin{flalign}
v_0 &= r_0 \, \Omega_K(r_0) \\
\rho_0 &=  M_\odot / r_0^3\\
P_0 &= \rho_0 v_0^2\\
B_0 &= \sqrt{4\pi \, \rho_0 v_0^2}
\end{flalign}
with $r_0 = 1 \, \mathrm{au}$ being the unit length.
For all simulations including X-ray heating, we applied a fiducial value of $L_X = 
2 \cdot 10^{30} \, \mathrm{erg}/\mathrm{s}$ for the X-ray luminosity from the 
central star.
\\
All calculations were performed on a 2D spherical grid in PLUTO, including the 
azimuthal components of the velocity and magnetic field. Relevant parameters and 
all simulation runs are summarized in Table \ref{tab:simulations}. The parameter 
$\beta_0$ refers to the initial plasma beta at the mid plane. The simulation 
time is represented in units of the orbital time scale at $r = r_0$. We chose the 
presented parameter range in $\beta_0$ to cover the regime of a significant 
magnetically driven wind, which has been found to emerge for $\beta_0 \approx 
10^5$ in previous studies \citep{Gressel2015b, Bai2017b}. By increasing $\beta_0$ by several orders of 
magnitude, the magnetically driven wind is expected to vanish, and the transition 
toward a photoevaporation dominated flow should occur. \\
Except for run X-bn-h, the polar angle spanned from $ \theta = 0.005$ to $\theta = \pi 
- 0.005$ in order to avoid numerical difficulties at the rotation axis. In the 
polar direction, the grid was stretched with an overall ratio of $\approx 4$ to 
provide sufficient resolution throughout the disk atmosphere while saving 
computational power in the less critical upper wind region.  In other words, the 
stretched grid can be expressed as follows: $\Delta \theta (\theta \approx 0) = 
4 \, \Delta \theta (\theta \approx \pi / 2),$ with $\Delta \theta$ being the 
extent of the grid cell in polar direction.
The only exception was the simulation run X-b6-mri, where a uniformly space grid was chosen in order to test the influence of resolved MRI-modes with respect to the global solution.
 To cover the dynamical range in 
radial direction, a logarithmic grid was chosen. With this strategy, a 
resolution of $\approx 25$ cells in polar and $\approx 6$ cells in radial 
direction per pressure scale height at $2\,\mathrm{au}$ in the mid-plane was 
achieved (except for X-b6-mri).\\
In order to obtain an outward-bent magnetic field topology, the bending parameter $m$ in Eq. 
\ref{eq:vector_potential} was set to 0.4. Because the plasma beta is relatively 
large, we do not expect significant magnetocentrifugal effects and the influence 
of variation in $m$ should be small. \\
For all simulations, the enclosed disk mass was $\approx 0.05 M_\odot$ in the 
domain. We did not apply an exponential cutoff to the density power law for 
stability reasons. The total disk mass is thus equivalent to a disk with a hard 
cutoff at the simulation boundary.
Furthermore, the gravitational potential caused by the central star was set to 
solar conditions. 

\section{Results} \label{sec:results}
Before employing the whole model including MHD terms, the photoevaporation 
approach was tested and compared to previous work. For this purpose, 
hydrodynamic simulations were carried out to study the wind topology and 
mass-loss rates. \\
The hydrodynamic stability of the disk setup was tested up to $10^4 \Omega^{-1}$  with EUV and X-ray heating switched off and no magnetic fields. No wind or outflow forms in these conditions.
\subsection{Photoevaporation}
An approximately stationary flow configuration is reached within the simulation 
time frame (in addition to fluctuations in the upper disk atmospheres beyond $\approx 
60\, \mathrm{au}$). Time-averaged results from 555 to 714 years of run 
X-bn-h are shown in Fig. \ref{fig:photoev_avg}.
When we consider the topology of the poloidal velocity streamlines, the velocity 
field resembles a radial field and exhibits no turbulent features toward the 
outer radial simulation boundary. Based on examining the sonic surface, visualized as 
a white line in the two plots, we state that the flow becomes 
supersonic, close to the wind-launching front. \\
Similar properties have been observed 
by \cite{Owen2010}. Slightly different wind-loss rates can be caused 
by a difference in the underlying disk model and the chosen resolution. 
Additionally, no exponential cutoff was applied in the simulations here. Because 
the runs, including two hemispheres, were restricted to an outer radius of 
$60\,\mathrm{au}$, the wind-loss rates are expected to be lower but 
the difference should be small because most of the wind mass-flux 
occurs at smaller radii.
When the results of simulation X-bn-h are averaged, the total mass-loss rate produced by 
the wind flow is $(1.38 \pm 0.06) \cdot 10^{-8} M_\odot / \mathrm{yr}$, which 
agrees well with the rates of \cite{Owen2010} $(1.4 \cdot 10^{-8} M_\odot / 
\mathrm{yr})$. \\
\begin{figure}[h!]
\centering
\includegraphics[width=0.9\linewidth]{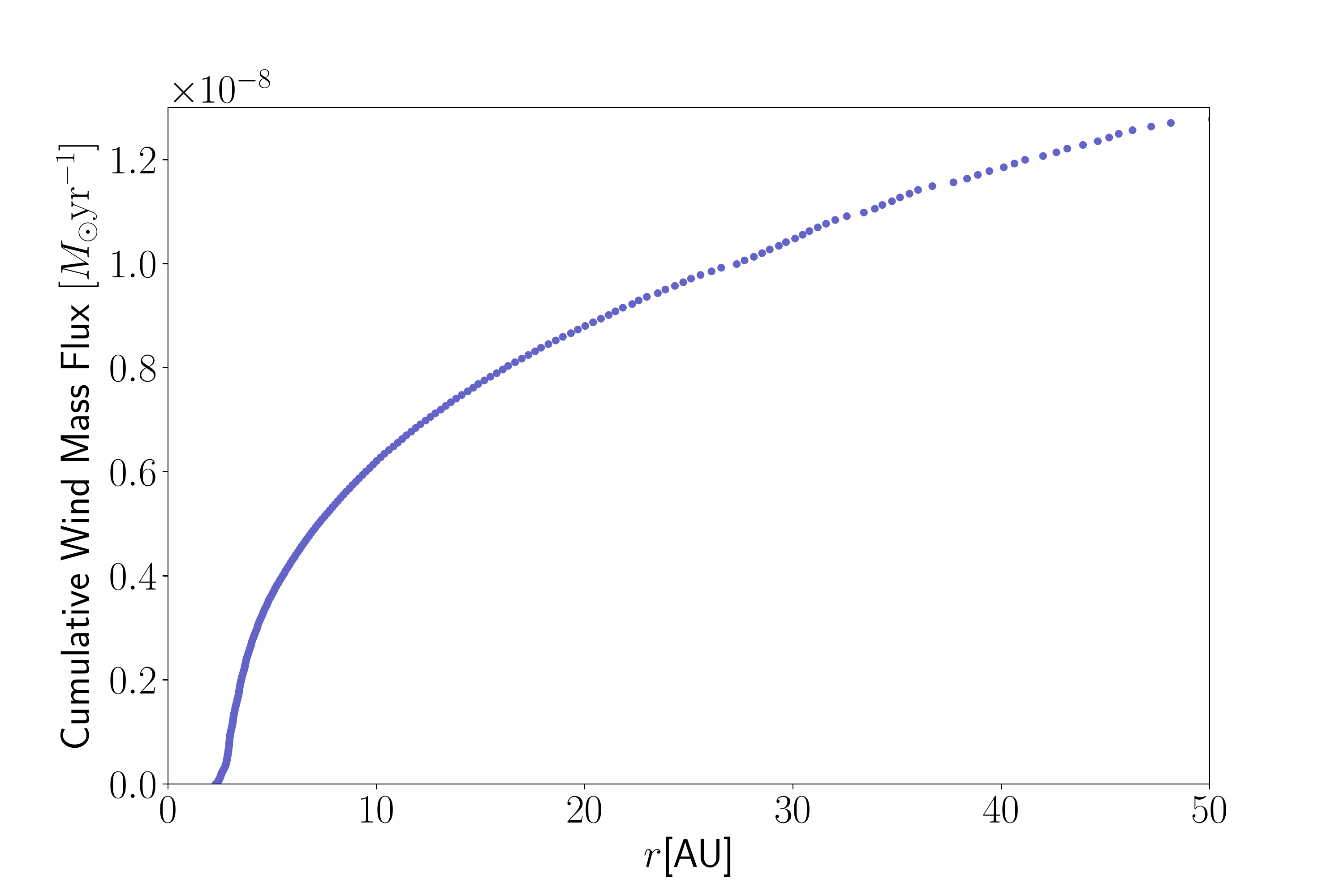}
\caption{Cumulative wind mass-loss rate in dependence of the radius in the mid-plane, resulting from time-averaged flows from $3.5 \cdot 10^4 \Omega^{-1}$ to $ 
4.5 \cdot 10^4 \Omega^{-1} $. It shows that approximately half of  the mass loss 
occurs within $15\,\mathrm{au}$.}
\label{fig:cumulativeLoss}
\end{figure}
\begin{figure}[h!]
\centering
\includegraphics[width=0.9\linewidth]{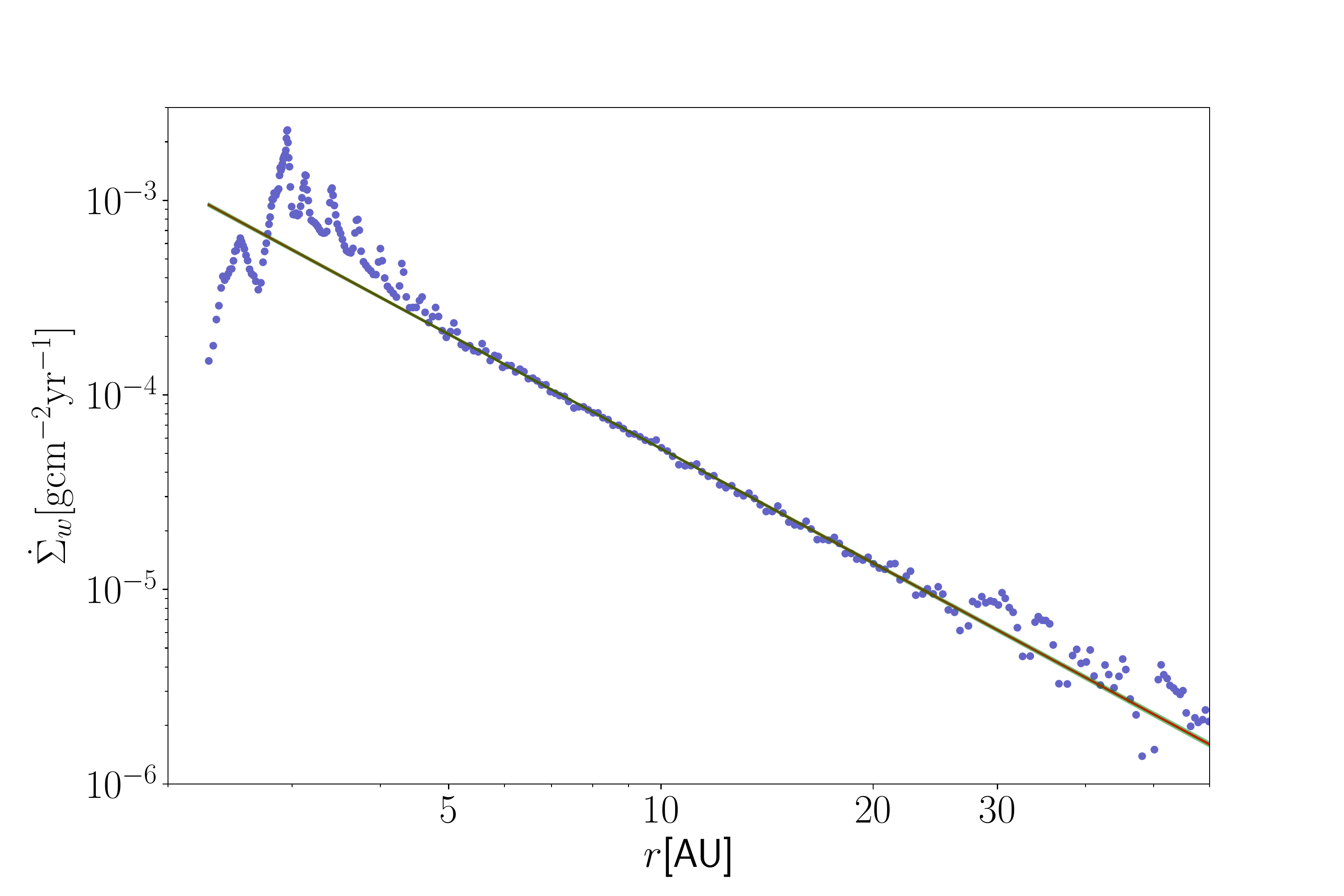}
\caption{Column density loss in g / $\mathrm{cm}^2$ and years with a fitted 
power law. Within $5\,\mathrm{au,}$ the wind rate clearly diverges from the power 
law, which is also true in the region near the outer simulation boundary. The two green 
lines represent the one-sigma margin of the fit result.}
\label{fig:columnLoss}
\end{figure}
\begin{figure*}[ht]

\subfloat[]{ \label{photoev_vfield}
        \includegraphics[width=0.5\textwidth]{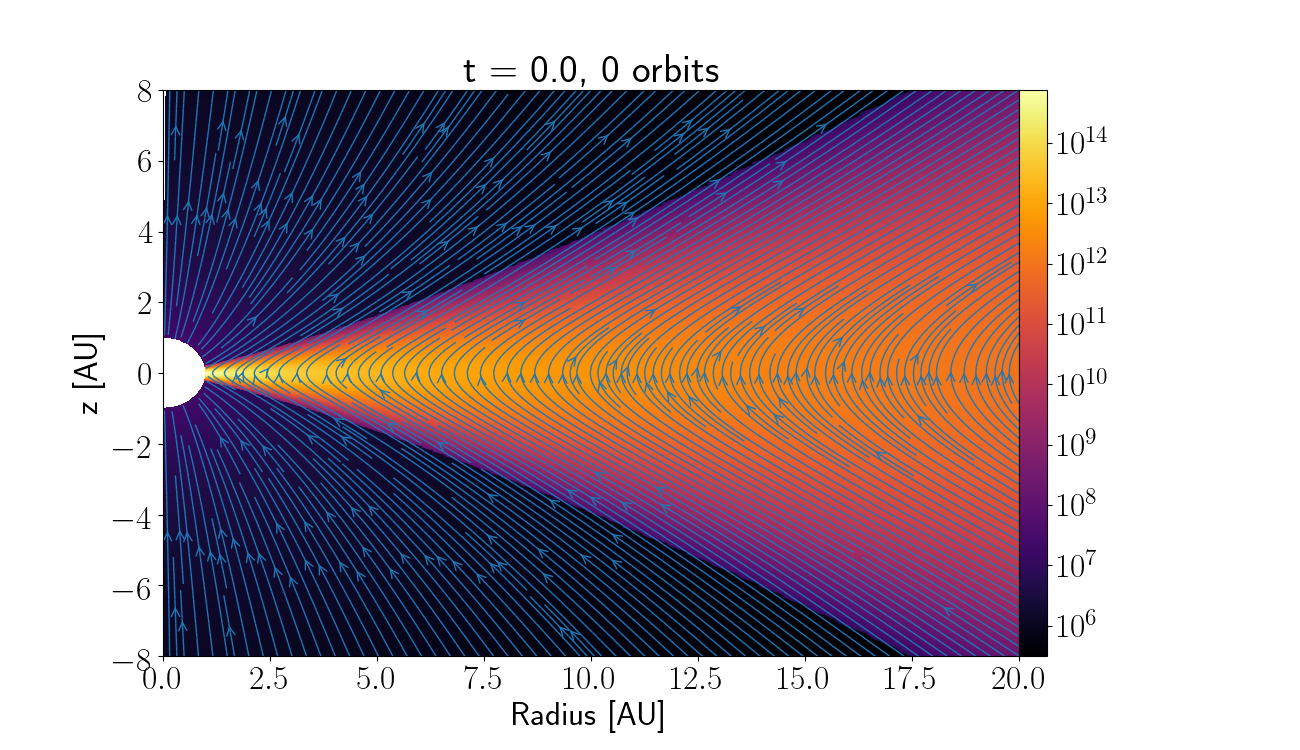}
}
\subfloat[]{ \label{photoev_vfield}
  \includegraphics[width=0.5\textwidth]{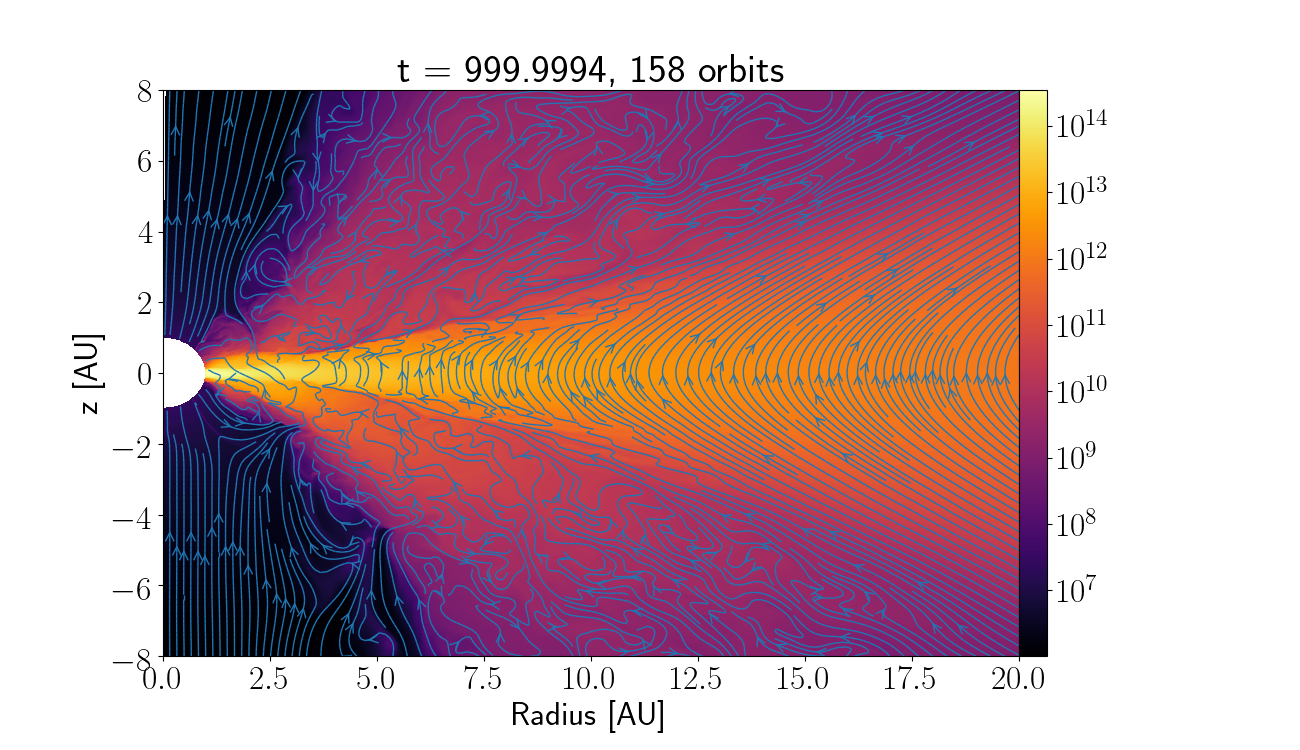}
}
\vskip\baselineskip
\subfloat[]{ \label{photoev_vfield}
        \includegraphics[width=0.5\textwidth]{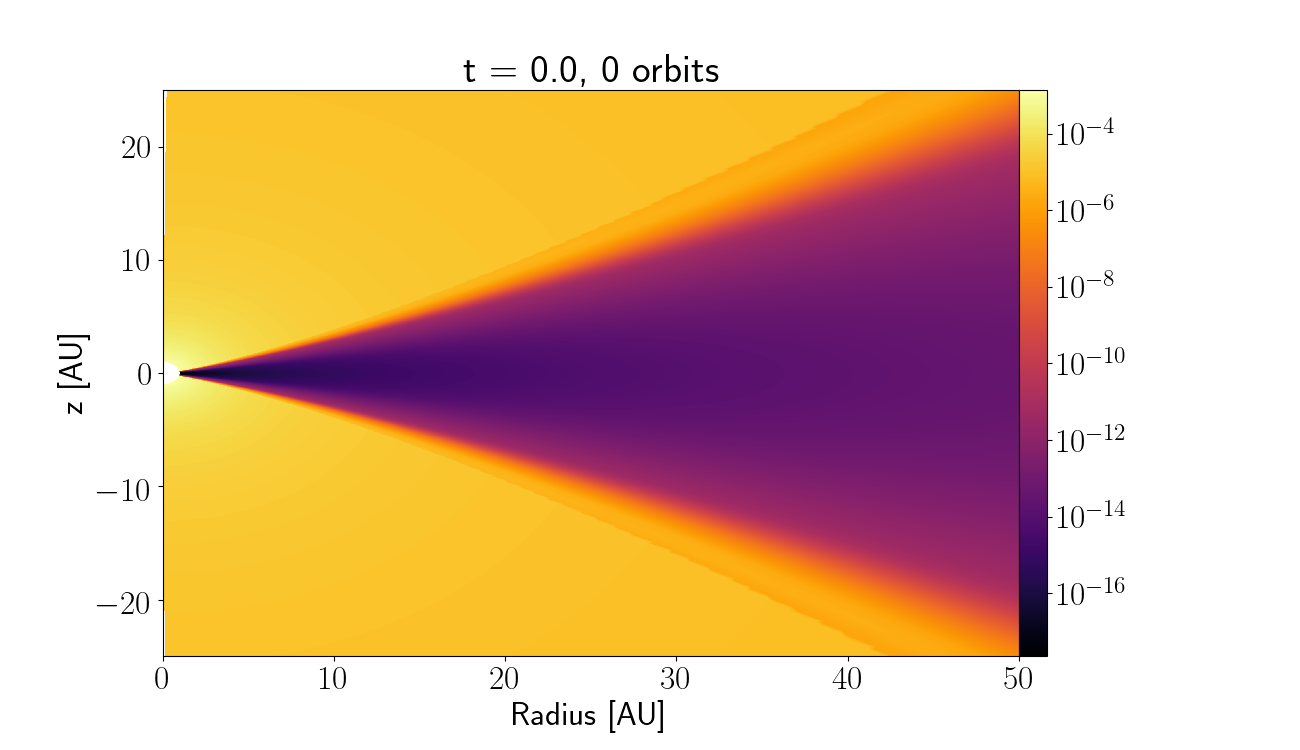}
}
\subfloat[]{ \label{photoev_vfield}
        \includegraphics[width=0.5\textwidth]{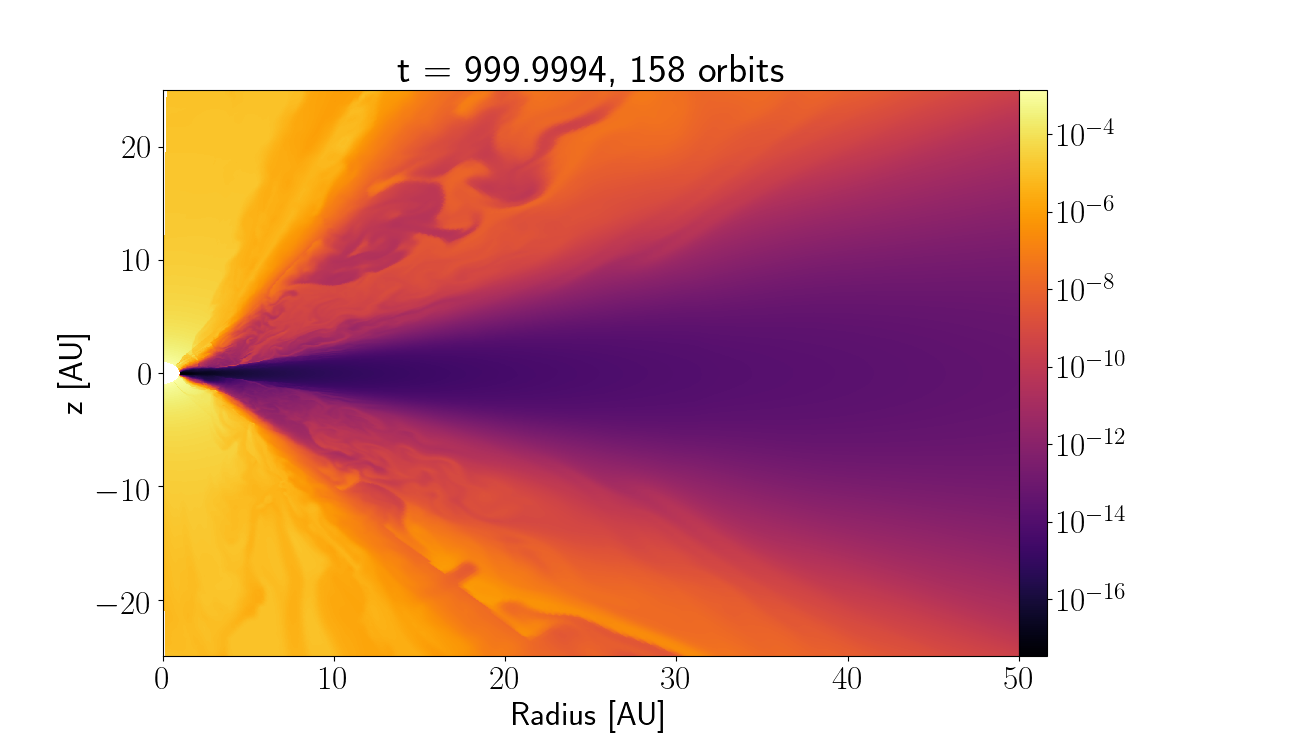}
}

\caption{Panels (a) and (b) visualize the number density in $\mathrm{cm}^{-3}$ and the magnetic field lines of 
simulation X-b5  in the initial configuration and after an 
evolution of 158 inner orbits at $r = r_0$ ($\approx 1.8$ orbits at 20 au), where a strong wind emerges. A slightly asymmetric 
flow forms in the inner region. 
Panels (c) and (d) depict the ionization fraction $x_e$ in 
the disk. The snapshot after 158 orbits at $r = r_0$ visualizes the reduced ionization 
fraction due to the thick wind flow.}
\label{fig:xb5_field}
\end{figure*}
Fig. \ref{fig:cumulativeLoss} shows the cumulative mass-loss rate versus the 
cylindrical radius $r$. The method here is to trace down the stream lines of the 
time-averaged flow to the wind-launching front and evaluate the radius at 
that point. The result indicates a very smooth profile, and it is clearly visible 
that within $15\,\mathrm{au,}$ half of the mass loss occurs. \\
 In Fig. \ref{fig:columnLoss} we plot the vertically integrated column density loss. Small errors in the estimation of the wind-launching front are possible because the gas trajectories do not necessarily stop at the ionization front, 
and it is challenging to exactly trace the origin computationally. For radii 
smaller than $5\,\mathrm{au,}$ the losses diverge from the approximate power law. 
In the outer part, deviations could originate from fluctuations near the 
boundary because the disk is observed to fail to reach static equilibrium there. 
For most of the simulated region, a power law of the following form can 
be fit: 
\begin{equation}
\dot{\Sigma}_w (r) = 4.76 \cdot 10^{-3} \left( \frac{r}{\mathrm{au}} 
\right)^{-1.95} \frac{\mathrm{g}}{\mathrm{cm}^2 \mathrm{yr}}
.\end{equation}

\subsection{MHD simulations}

When magnetic fields are introduced into the simulations, a variety of complex 
phenomena are added that can drastically alter the topology of the emergent flow structure. 
To determine the influence of MHD effects, a study with different values of the plasma parameter 
ranging from $10^5$ to $10^{10}$ was performed (see Table 
\ref{tab:simulations}). In the following section we examine the mass flux 
carried by the wind sand flow structure, and we study the evolution of the magnetic field.  
\\
\subsubsection{Wind flow} \label{sec:wind_accretion}
The magnetic field topology of run X-b5 is shown in the initial 
configuration and after an evolution of 158 inner orbits in Fig. 
\ref{fig:xb5_field}. Clearly, the magnetic field lines are bent outward and have 
an inclination of more than $30 \degree$ with respect to the rotation axis.
In the latter snapshot, a slightly irregular outflow is visible. In the inner 
region of the disk an asymmetric distribution of gas develops over time due to 
the increasingly wound up toroidal magnetic field toward the mid-plane. \\
The disk remains mostly laminar in the lower atmosphere, whereas radial 
perturbations of the magnetic field form in the upper layers. A development of 
the MRI cannot be excluded, but in a 2.5D framework, the saturation of the 
MRI cannot be accurately followed. 
A more detailed discussion and the impact of the numerical resolution of the MRI modes on the wind solution is given in section \ref{sec:transition}.
\begin{figure*}[t]

\subfloat[]{ \label{photoev_vfield}
        \includegraphics[width=0.5\textwidth]{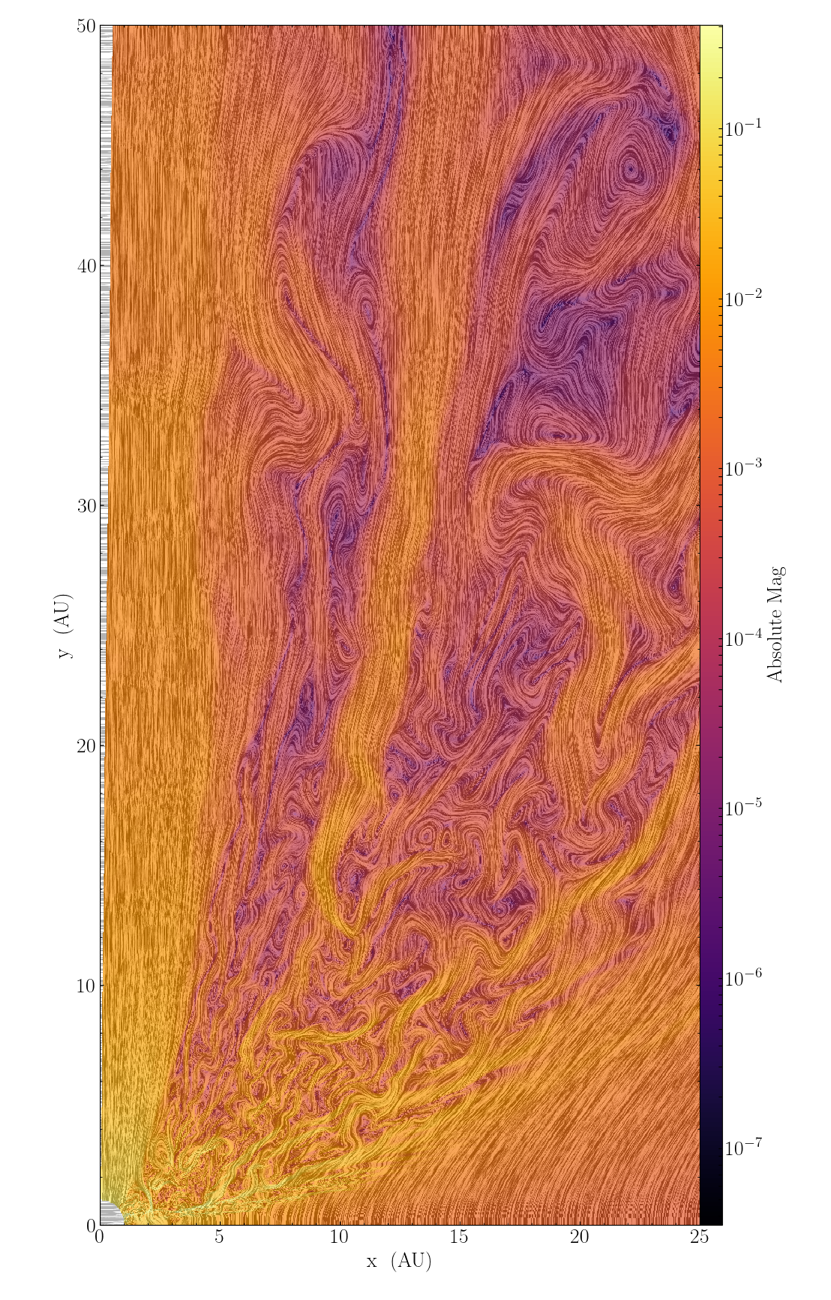}
}
\subfloat[]{ \label{photoev_vfield}
        \includegraphics[width=0.5\textwidth]{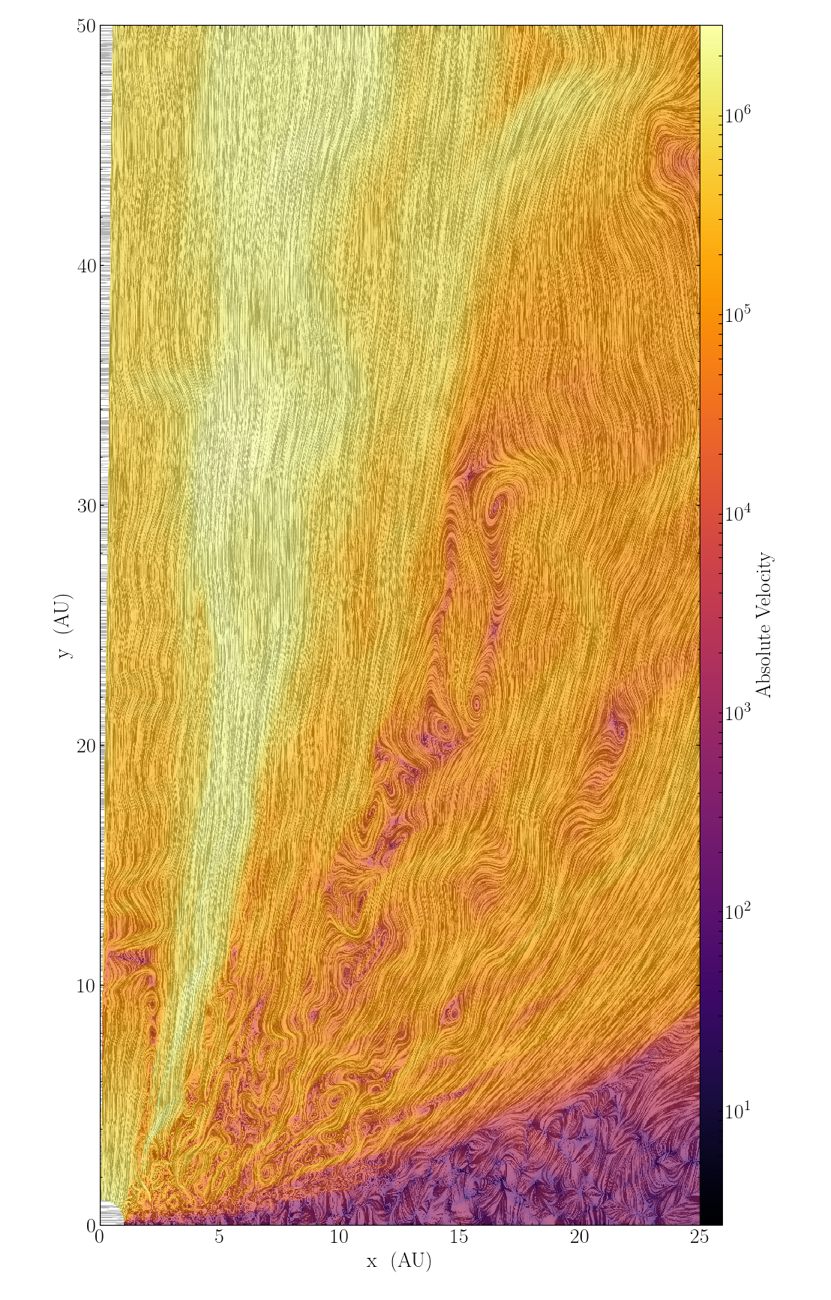}
}
\caption{On the left the magnetic field structure (LIC) is visualized for 
simulation X-b5 after 158 years. Only the upper hemisphere (both are simulated) is shown to highlight a larger portion of the wind structure. The color map represents the absolute poloidal 
magnetic flux density in Gauss. On the right-hand side the velocity profile is 
depicted, including the color as the absolute velocity in cgs units. The wind is 
not in a steady state, as the irregular magnetic field lines and stream lines indicate.}
\label{fig:xb5_lic}
\end{figure*}

\begin{table}[t]
\caption{Wind and accretion rates}
\centering
\begin{tabular}{lcccccc}
\hline
Run & $\beta$ & Wind flux [$M_{\odot} \mathrm{yr}^{-1}$] & Accr. rate \ 
[$M_{\odot} \mathrm{yr}^{-1}$]\\
\hline
\hline
X-b5  & $10^5$ & $(5.93 \pm 2.97) \cdot 10^{-7}$ & $(1.11 \pm 0.31) \cdot 10^{-7}$\\
X-b6  & $10^6$ & $(5.05 \pm 1.98) \cdot 10^{-8}$ & $(1.32 \pm 0.93) \cdot 10^{-8}$\\
X-b7 & $10^7$ & $(1.15 \pm 0.49) \cdot 10^{-8}$ & $(1.25 \pm 6.20) \cdot 10^{-9}$ \\
X-b8 & $10^8$ & $(8.40 \pm 1.66) \cdot 10^{-9}$ & $(0.23 \pm 6.25) \cdot 10^{-9}$ \\
X-b9 & $10^9$ & $(8.86 \pm 0.74) \cdot 10^{-9}$ & $(0.23 \pm 6.19) \cdot 10^{-9}$\\
X-b10 & $10^{10}$ & $(8.78 \pm 0.67) \cdot 10^{-9}$ & $(0.16 \pm 6.17) \cdot 10^{-9}$ \\ 
X-bn & \ldots & $(9.66 \pm 0.44) \cdot 10^{-9}$ & \ldots\\ 
\hline
\end{tabular}
\tablefoot{Wind mass-loss rates and accretion rates of all simulations including 
magnetic fields and X-ray photoevaporation. The rate at the bottom corresponds 
to the hydrodynamical simulation X-bn without magnetic effects.} 
\label{tab:windrates}
\end{table}

\begin{figure}[h]
\centering
\includegraphics[width=\linewidth]{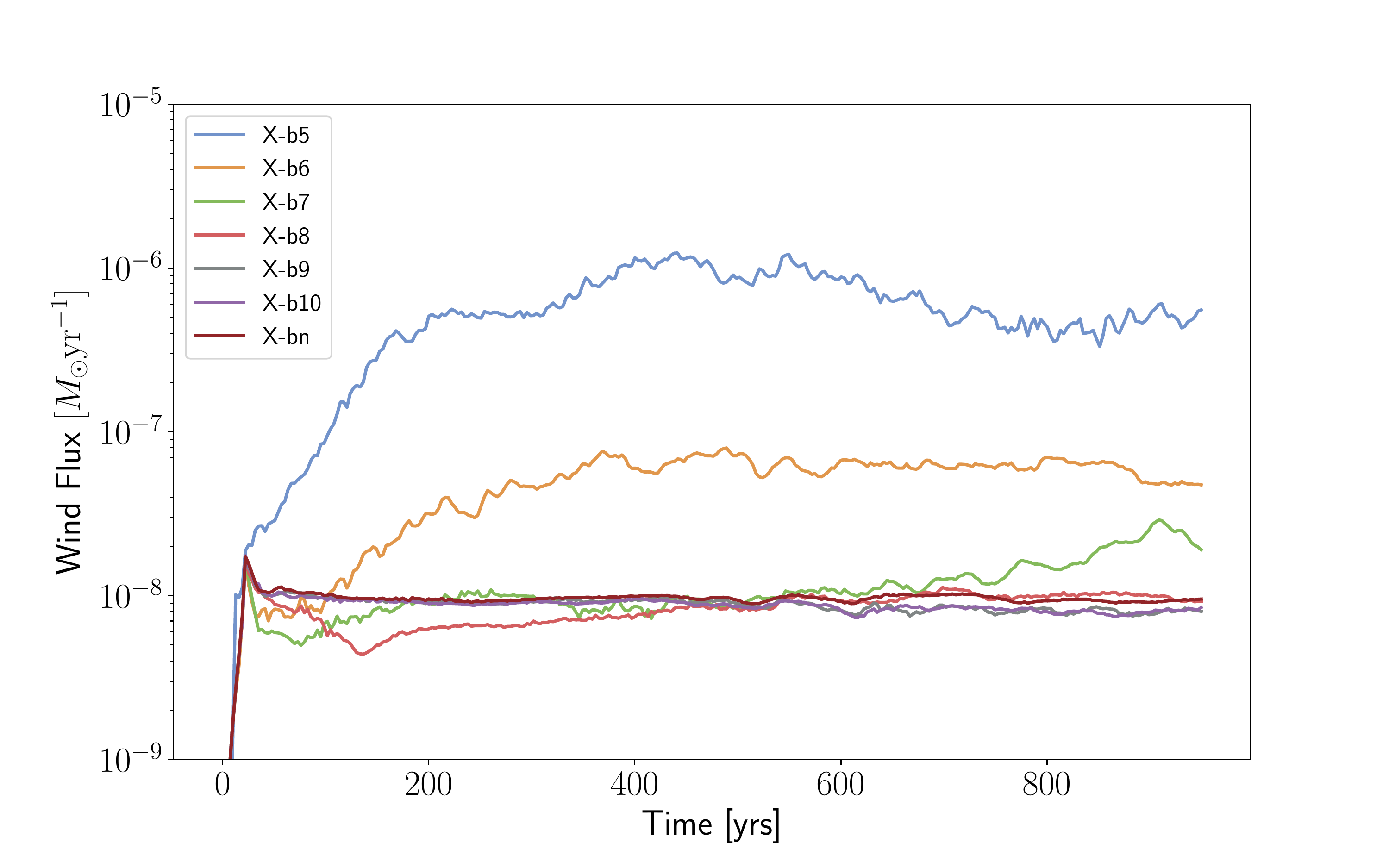}
\caption{Wind mass fluxes in solar masses per year for various 
simulation runs. For a plasma beta higher or lower than $10^7$, wind rates of about 
one order of magnitude higher than the photoevaporation rates are observed. For 
a plasma beta larger than $10^7$, lower wind rates emerge, which is an indirect 
result of radiation shadowing that is caused by slight magnetically induced turbulence in 
the inner disk region.}
\label{fig:windfluxes}
\end{figure}

\begin{figure*}[t]
\centering
\includegraphics[width=0.9\linewidth]{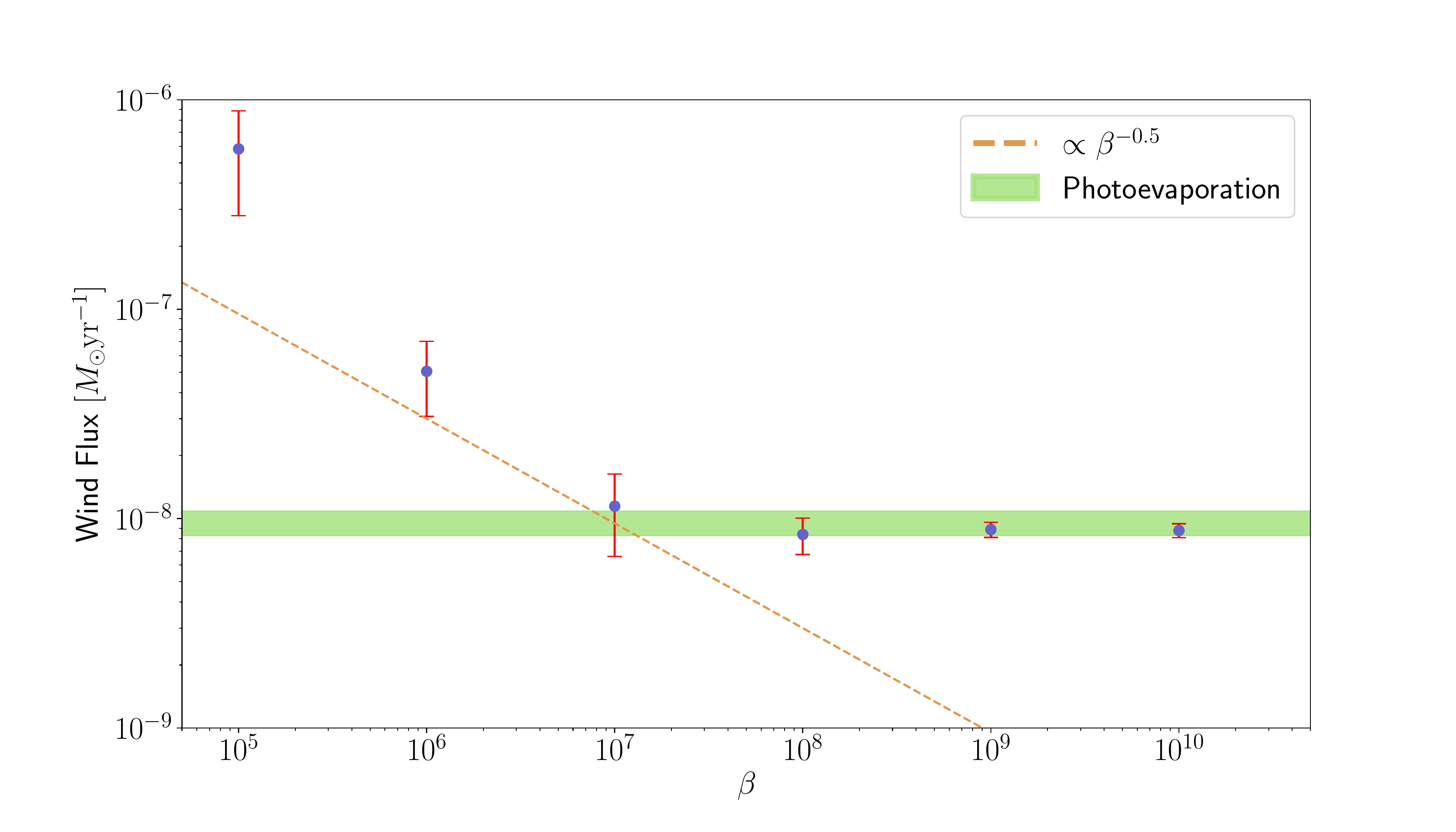}
\caption{Time-averaged wind fluxes vs. the plasma parameter 
$\beta$. The errors are estimated through the standard deviation of the time -averaged sample. For stronger fields, the rates approximately follow the same 
power law as the dashed line, which is based on results of \cite{Bethune2017}. 
For weaker fields, the photoevaporation begins to dominate, and the data points 
at $\beta > 10^7$ indicate the influence of radiation shielding in the inner 
region, which results in lower mass fluxes overall.}
\label{fig:windfluxes_errorbar}
\end{figure*}

\begin{figure*}[t]
\subfloat[]{ \label{photoev_vfield}
        \includegraphics[width=0.5\textwidth]{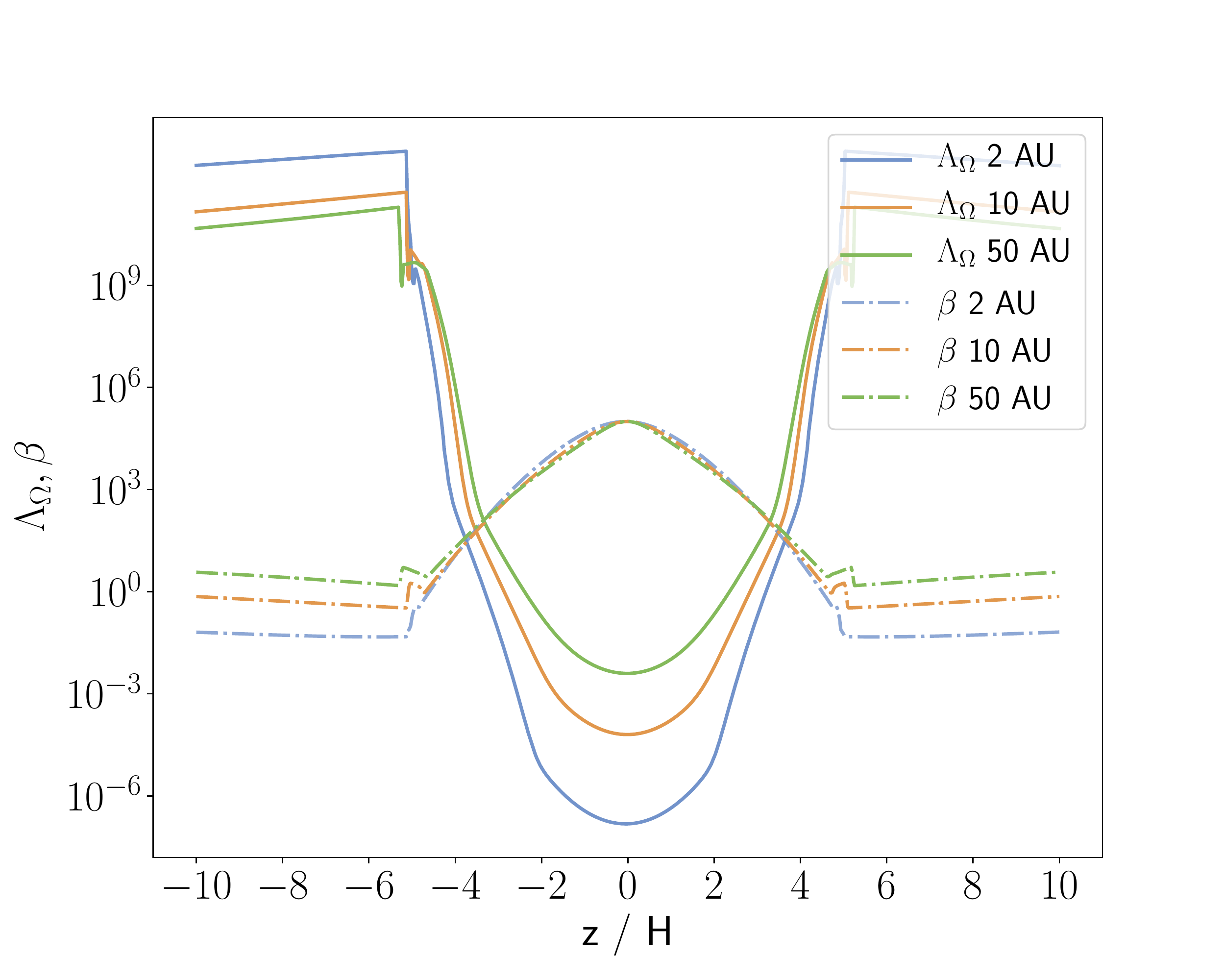}
}
\subfloat[]{ \label{photoev_vfield}
        \includegraphics[width=0.5\textwidth]{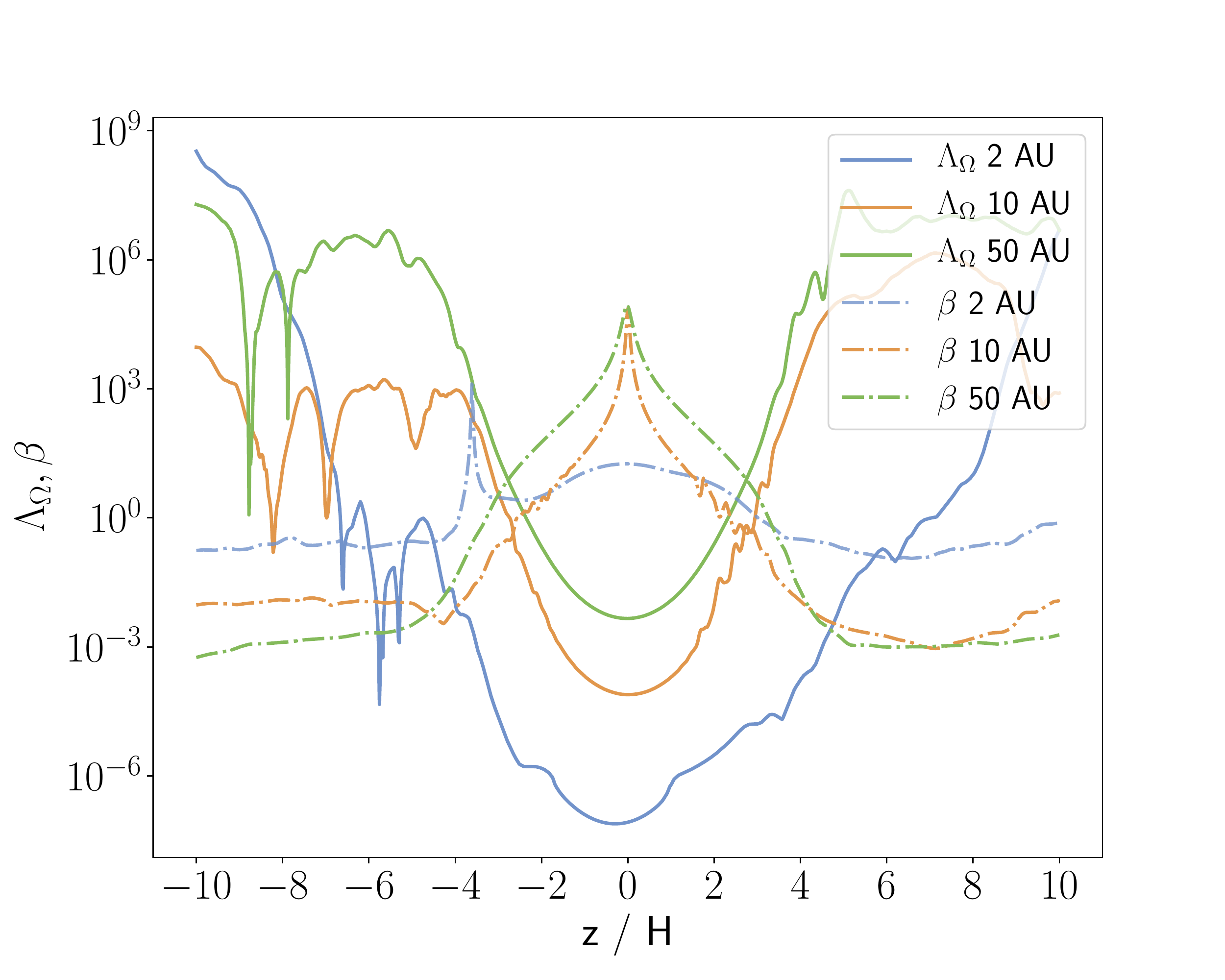}
}
\caption{ Panels (a) and (b) show the vertical distribution of the Elsasser 
number $\Lambda_\Omega$ and the plasma beta for different radii at an 
initial $\beta = 10^5$ in the mid-plane. The second panel indicates a time-
averaged evolved state.}
\label{fig:xb5_elsasser}
\end{figure*}
In Fig. \ref{fig:xb5_lic} the wind structure in the upper hemisphere is 
displayed for simulation X-b5. Within a cylindrical radius of $r \approx 5\, 
\mathrm{au}$, the magnetic and velocity field roughly coincide, whereas for 
outer radii, significant differences in the flow structure are visible, indicating that the flow is not in steady state.
Fig. 
\ref{fig:xb5_field} (c) and (d) show the ionization fraction for the 
initial snapshot and after more than 150 orbits, respectively. The thick wind 
layer in the inner part of the disk partly blocks the radially penetrating 
radiation. For a wide range of the outer disk, the ionization fraction is 
therefore significantly lower than the optically thin region. In the two figures, thermal 
ionization is not taken into account (the regions with $x_e > 10^{-4}$ would be 
approximately fully ionized because of the photoevaporation temperature 
prescription). \\
In Fig. \ref{fig:windfluxes} we show the evolution of the wind rate over time. 
The mass-loss rates generally increase with lower plasma beta. For $\beta = 
10^5$ the convergence to steady state is relatively slow because the increasing 
toroidal field alters the structure from the inner disk outward.
At about $\beta \approx 10^7$ , saturation into a mostly thermally driven flow is 
reached. For weaker magnetic fields, fluctuations are still present because the 
photoevaporative flow is sensitive to changes of the vertical extent of the 
inner disk. Chunks of gas moving to the upper layers inhibit the irradiation of 
outer parts of the disk and alter the effectiveness of the heating prescription.
\\
Fig. \ref{fig:windfluxes_errorbar} can be considered as the main result of this 
work. We show wind mass fluxes versus the initial plasma beta. The green 
region indicates the mass flux observed in the ideal photoevaporation run X-bn. 
The orange dashed line represents a mass-loss rate taken from the simulations of 
\cite{Bethune2017}, which scales as $3 \cdot 10^{-5} \beta^{-1/2} M_\odot 
\mathrm{yr}^{-1}$. A similar relation has also been observed in the shearing box 
simulations of \cite{Bai2013a}. 
Here, the measured mass-loss rates are higher than those obtained by 
B\'ethune (see Table \ref{tab:windrates} for numerical values). On the one hand, 
differences in the disk model can account for the discrepancy. On the other 
hand, ambipolar diffusion, dominating in the upper layers of the disk, would 
likely lower the mass-loss rates, as observed in \cite{Gressel2015b}. 
\\
For $\beta > 10^7$ the results clearly diverge from the fitted relation and 
approach a saturation point that corresponds to the photoevaporation rate 
without magnetic fields.
A magnetically driven wind cannot be launched at these initial field strengths, 
but the perturbations in the upper layers of the disk are still able to alter 
the thermal wind flow. These stochastic deviations of the mass-loss rate are 
visible in Fig. \ref{fig:windfluxes}. 
The photoevaporation rate is about 30\% lower than the result of X-bn-h because 
the simulation domain is smaller.

\subsubsection{Magnetic field evolution}

\begin{figure}[h]
\centering
\includegraphics[width=1.15\linewidth]{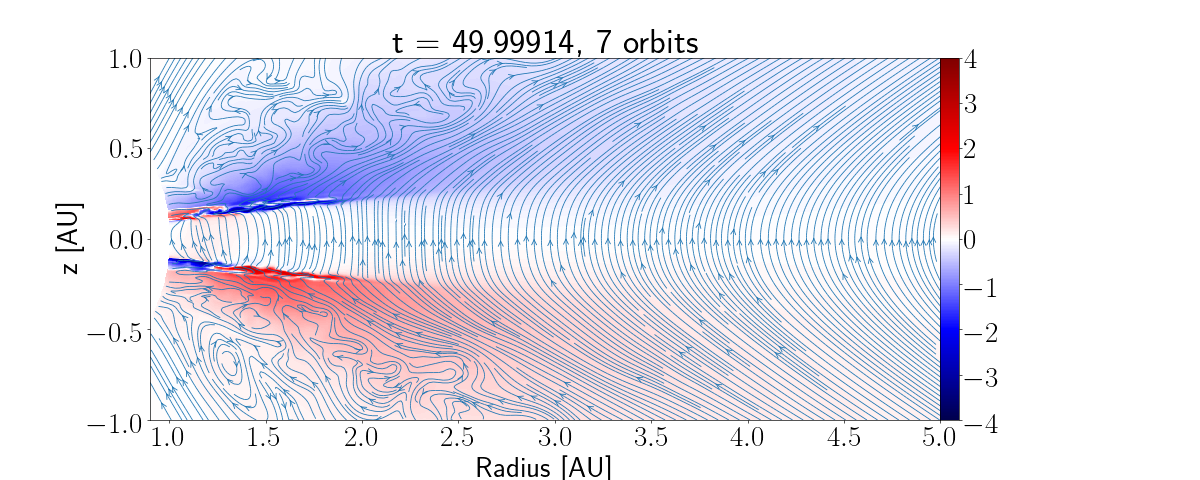}
\includegraphics[width=1.15\linewidth]{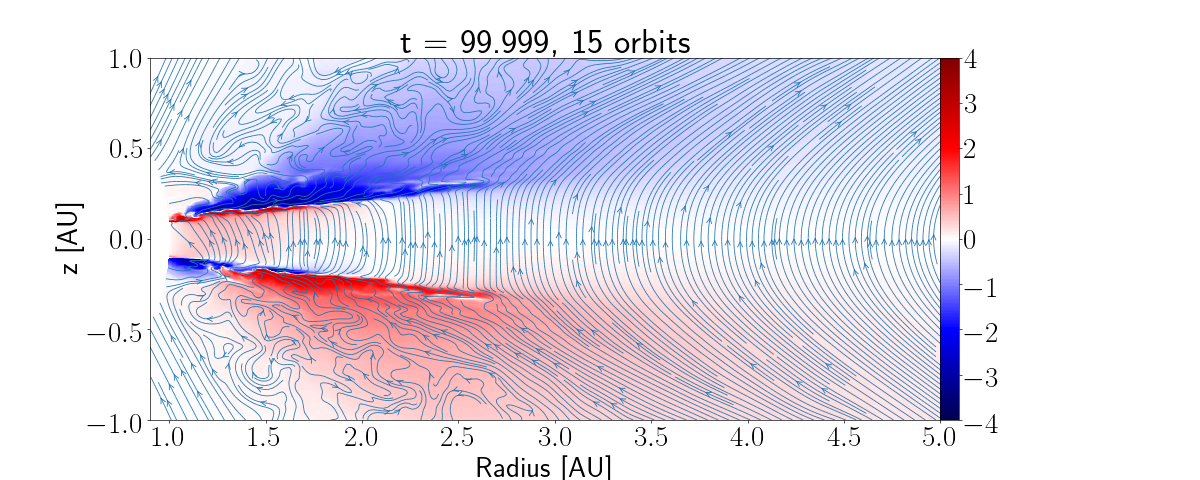}
\includegraphics[width=1.15\linewidth]{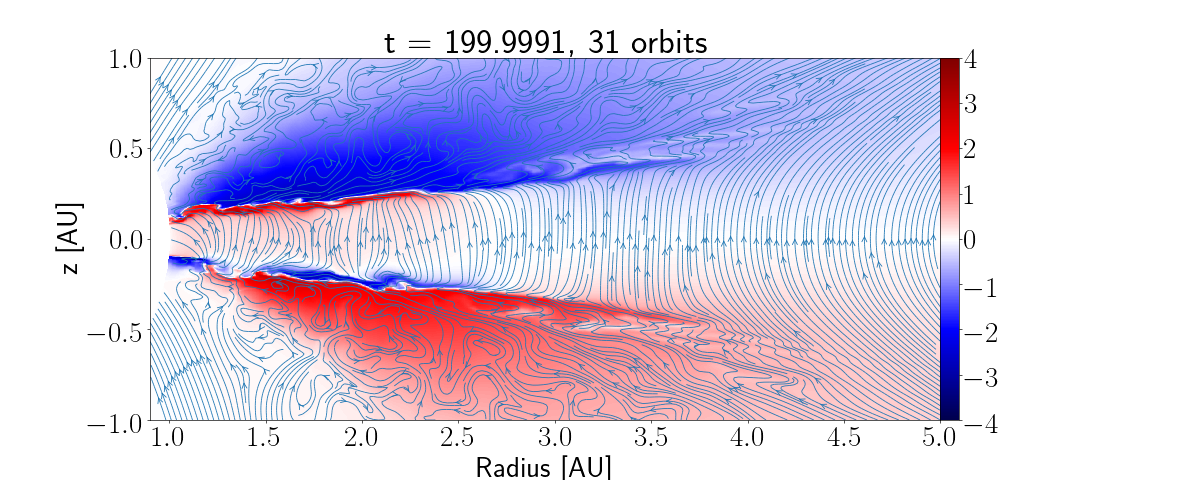}
\includegraphics[width=1.15\linewidth]{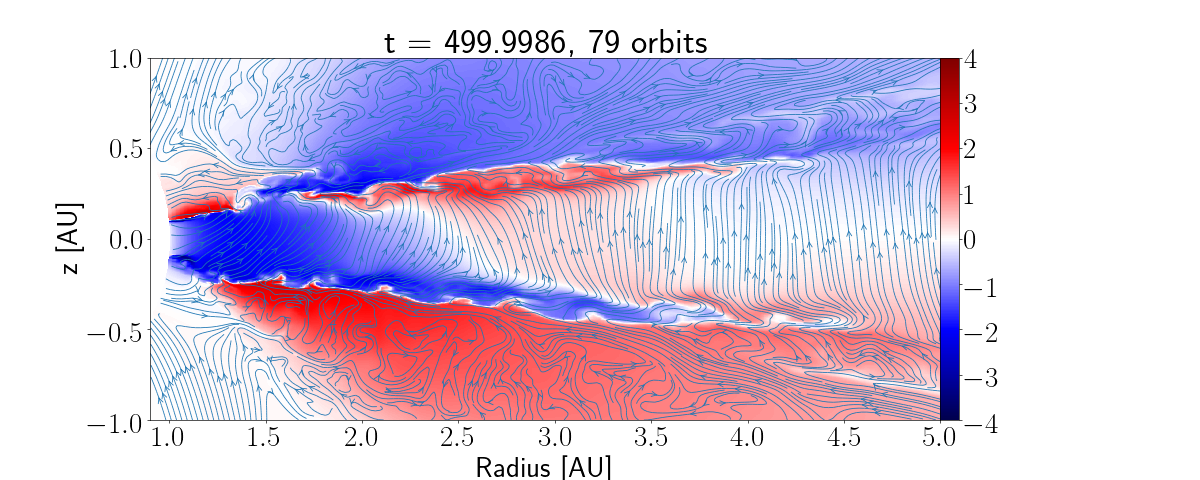}
\caption{Development of the toroidal magnetic field in the 
inner disk region in cgs units of simulation X-b5. The lines refer to the 
poloidal magnetic field lines.}
\label{fig:field_toroidal}
\end{figure}

In the following, we pursue a more detailed study of flow properties and magnetic field 
evolution. All figures and studies are based on simulation run X-b5. \\
The two diagrams (a) and (b) in fig. 
\ref{fig:xb5_elsasser} show the ohmic Elsasser number $\Lambda_\Omega$ and the 
local plasma beta versus the vertical height, scaled to the initial pressure 
scale height, for different radii. 
Within $\approx 3.5H,$ the Elsasser numbers are initially smaller than one, where 
MRI is expected to be suppressed by the ohmic diffusion, see \cite{Wardle2007}. The 
plasma beta decreases below unity in the upper layers because of the vertically 
exponential density and pressure distribution. For larger radii, the Elsasser 
numbers increase in the mid-plane because of lower densities and larger 
ionization fractions. In the time-averaged evolution in diagram (b), 
$\Lambda_\Omega$ decreases in run X-b5 at inner radii in the lower 
hemisphere where the asymmetric density structure builds up. The value of $\beta$
increases in the inner region because the toroidal magnetic field strength 
becomes amplified during its evolution. \\
Fig. \ref{fig:field_toroidal} displays the evolution of the toroidal magnetic 
field. 
Initially zero, the field winds up as a result of Keplerian motion and 
vertical shear in the layer farther away from the mid-plane.
The increasing toroidal field in the mid-plane causes the asymmetric polarity 
distribution in vertical direction, as also observed in \cite{Bethune2017}.
Slightly irregular flow structures and magnetic field lines are apparent. The 
origin may be attributed to MRI. A saturation cannot be achieved in 2.5D 
dimensions, and fully developed turbulence is absent in the simulations 
presented here.
In addition, experimenting with the boundary conditions, for instance, setting the 
toroidal magnetic field component to zero in the ghost cells of the inner radial 
boundary does not change the observed phenomena significantly. \\ 

\subsubsection{Flow analysis} \label{sec:flow_analysis}

\begin{figure}[h]
    \centering
    \includegraphics[width=\linewidth]{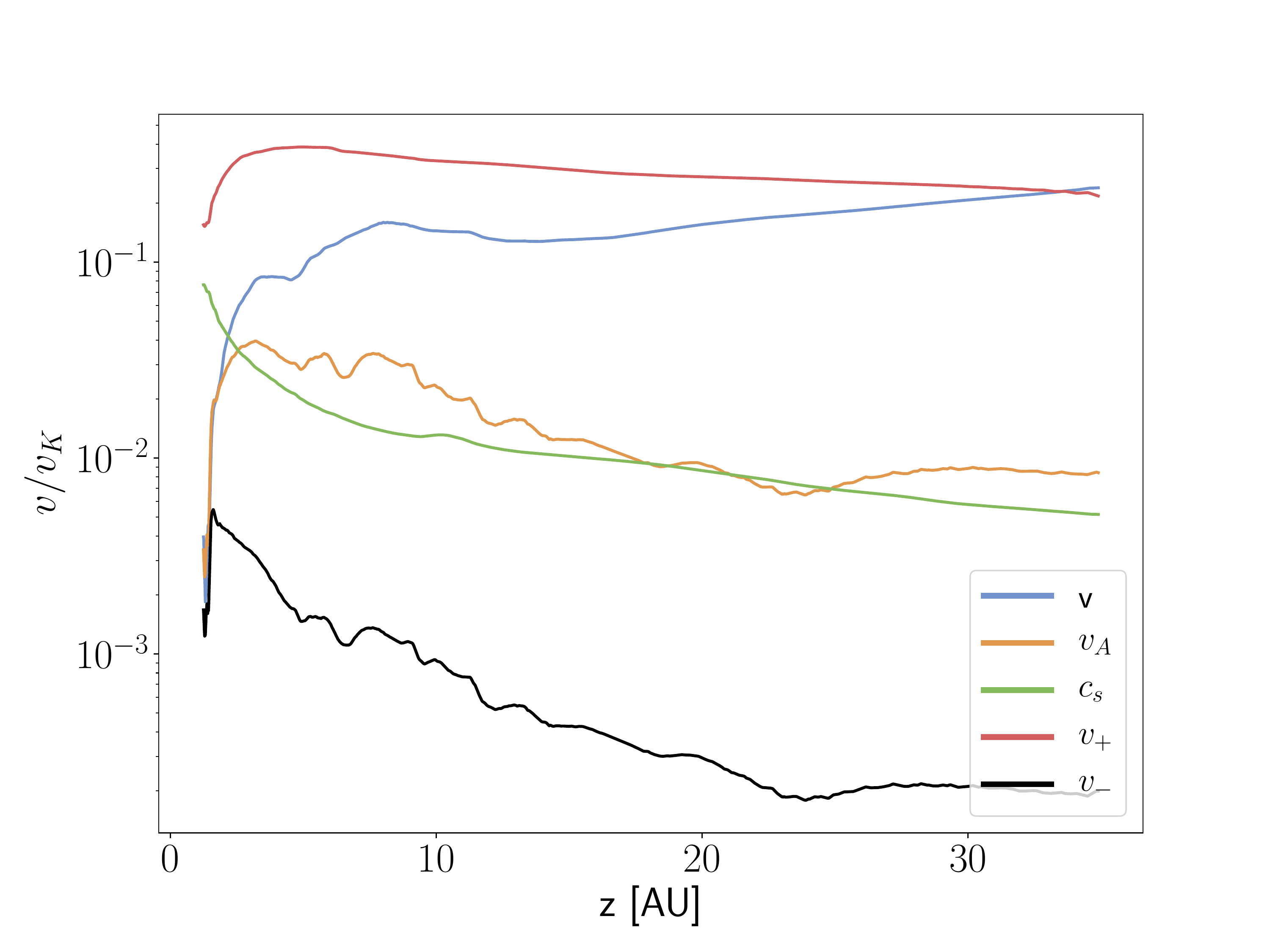}
    \caption[Poloidal velocities in X-b5]{Relevant poloidal velocities 
developing along a streamline in the time-averaged flow of X-b5. All velocities 
are normalized with the Keplerian velocity at the corresponding foot-point 
radius. Both the Alfv\'en point and the fast magnetosonic point are contained 
within the simulation domain. The blue line indicates the actual poloidal 
velocity, $v_A$ the local Alfv\'en speed, $c_s$ the sound speed, and $v_\pm$ the 
fast and slow 
%\LEt{please rephrase what the slash signifies here for clarity}
magnetosonic speed.}
    \label{fig:streamline_xb5}
    \includegraphics[width=\linewidth]{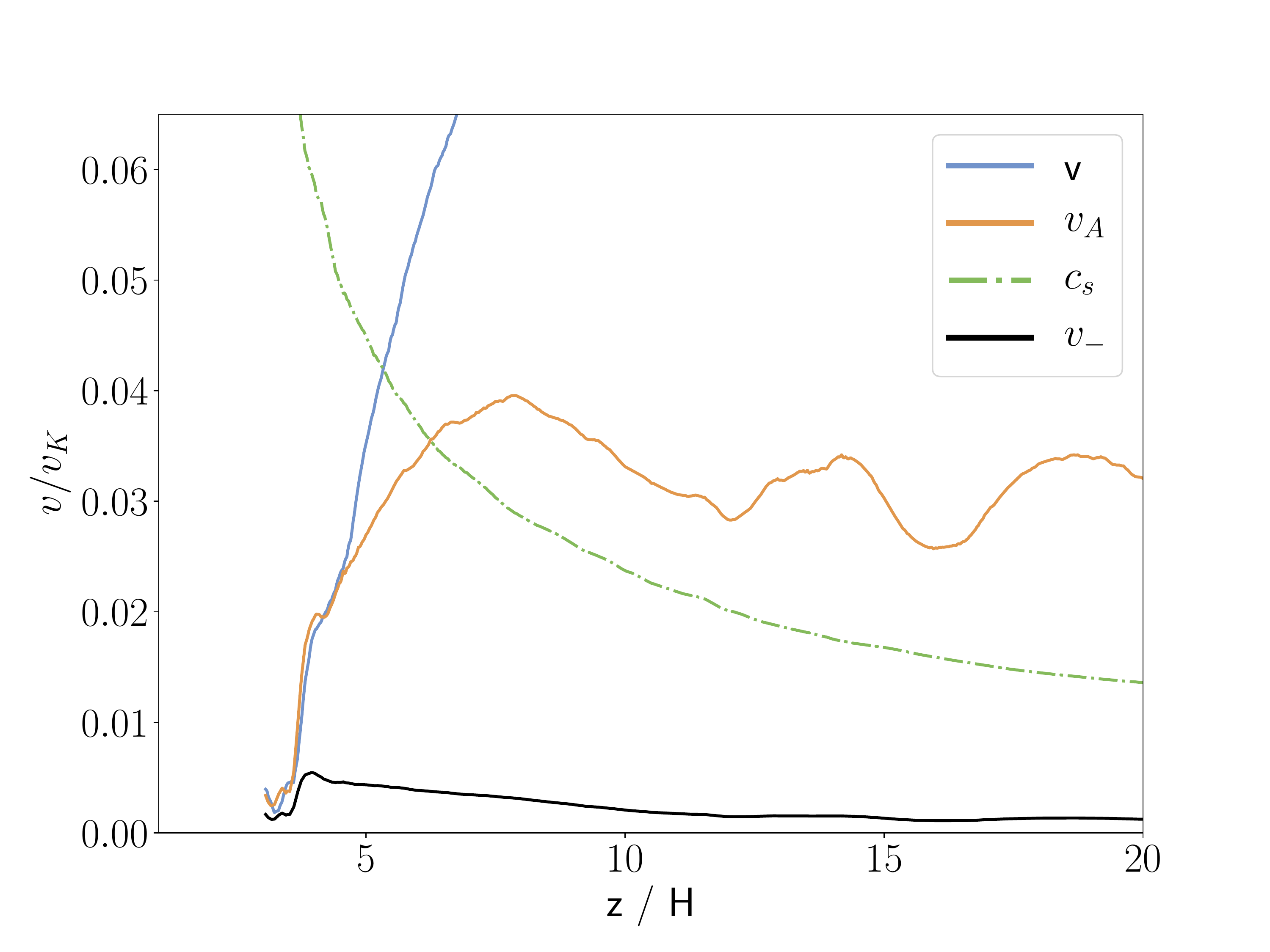}
    \caption[Poloidal velocities near the foot point]{Zoomed-in persepective 
of the same streamline as in Fig. \ref{fig:streamline_xb5}. This resolves the location 
of the Alfv\'en point, which lies relatively close to the wind-launching 
front.}\label{fig:streamline_xb5_zoom}
\end{figure}

To examine the wind properties, we traced various variables through a poloidal 
($r, z$) streamline in the flow. 
Again, run X-b5 serves as a reference model here.
The actual 3D trajectory of the gas differs from the poloidal streamline because the azimuthal velocity component is zero. 
In this context, the streamlines are only well defined in a truly steady-state 
flow. Nonetheless, we used the concept of streamlines to gain insight into the 
flow patterns and the contributing forces.\\
All the following plots were computed using the same integrated streamline in 
their respective simulations.
In Fig. \ref{fig:streamline_xb5} and Fig. \ref{fig:streamline_xb5_zoom} we show the 
poloidal velocities along a streamline in a time-averaged flow of X-b5, 
originating from $(r=6, z=1.3)\,\mathrm{au,}$  where $v_+$ and $v_-$ are 
the fast and slow mangetosonic velocities,
\begin{equation}
v_{\pm} = \frac{1}{2}  \left[ (c_s^2 + v_A^2)  \pm \sqrt{(c_s^2 + v_A^2)^2 - 4 
c_s^2 v_{Ap}^2} \right]
,\end{equation}
with the poloidal Alfv\'en velocity $v_\mathrm{Ap}$.
 The Alfv\'en point is reached close to the wind-launching front, as well as the 
sonic point. At $z \approx 28\, \mathrm{au,}$ the poloidal flow velocity passes 
through the fast magnetosonic point.
The definition of the wind-launching front for arbitrary radii is difficult in 
this case because the surface structure of the disk is rather chaotic. The 
streamline studied here merely acts as an example of the flow and indicates a 
rough position of relevant characteristic points along the wind flow.

\begin{figure}[h]
    \centering
    \includegraphics[width=\linewidth]{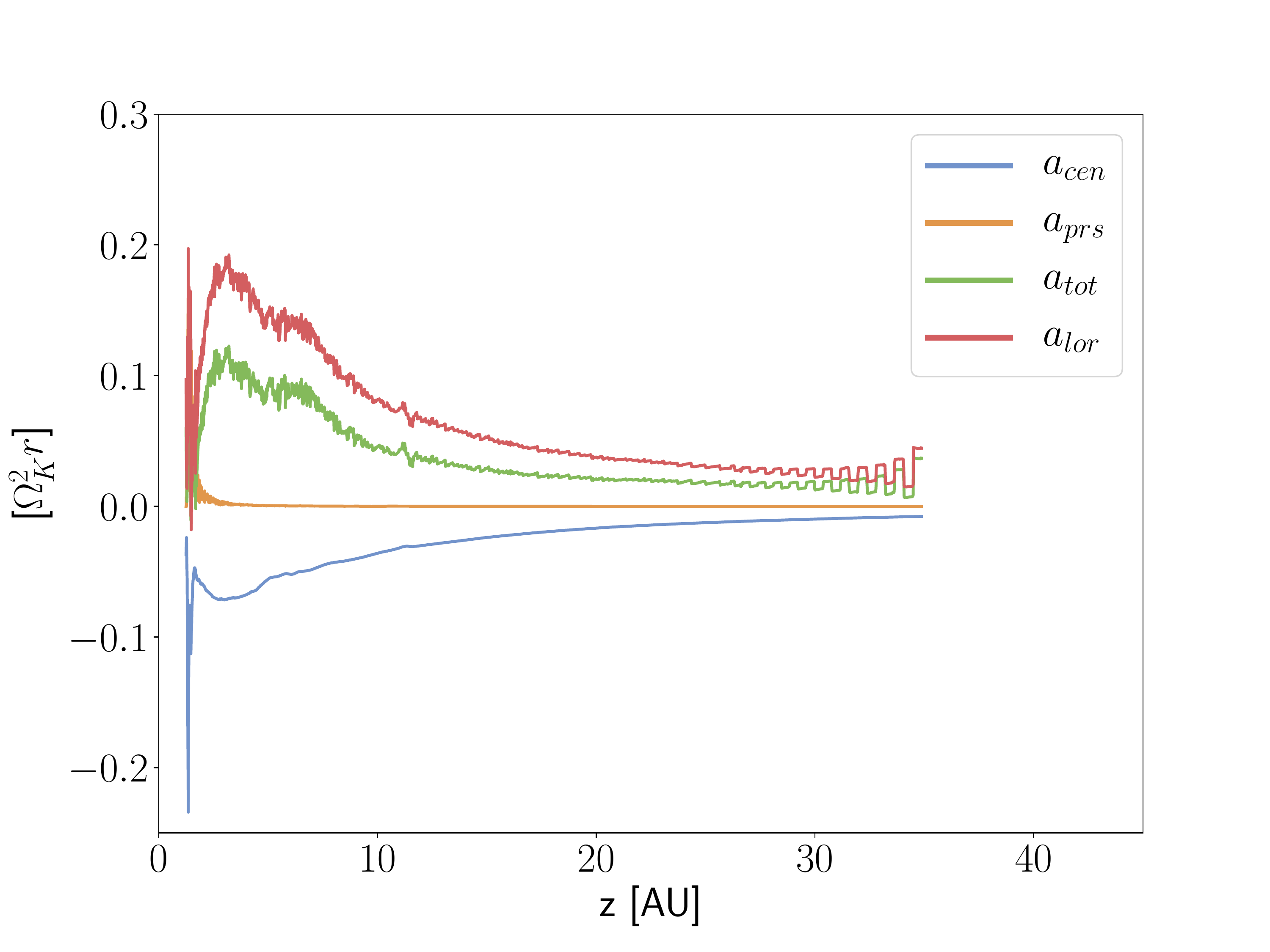}
    \caption{Decomposition of the contributing accelerations along a poloidal 
streamline for run Xb-5. Here, $a_\mathrm{cen}$ denotes the inertial 
acceleration, $a_\mathrm{prs}$ the acceleration due to the thermal pressure 
gradient, and $a_\mathrm{lor}$ the acceleration by the magnetic pressure 
gradient.}
    \label{fig:streamline_xb5_forces}
\end{figure}

\begin{figure}[h]
    \centering

    \includegraphics[width=\linewidth]{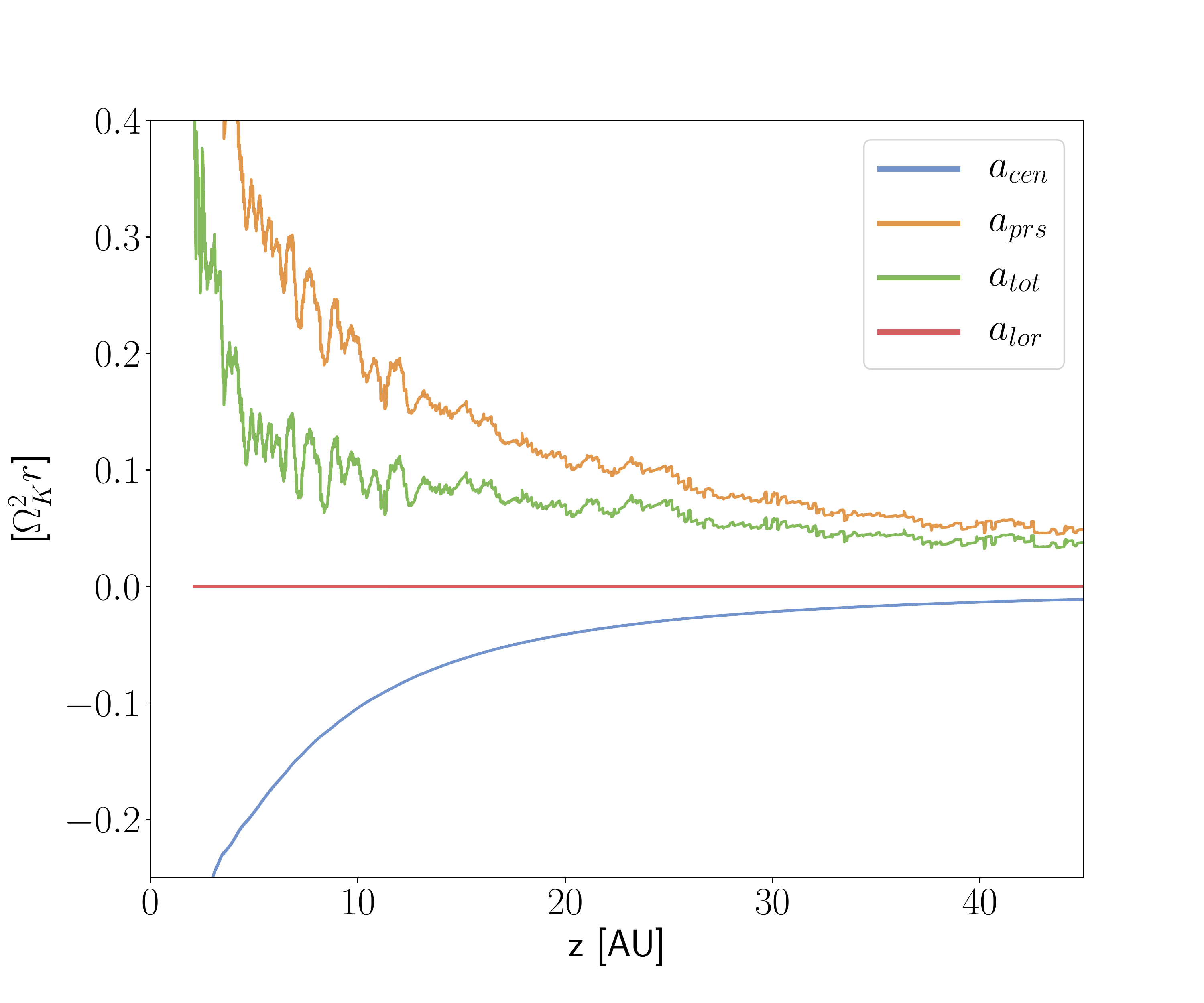}
    \caption{Decomposition of the contributing accelerations along a poloidal 
streamline for run Xb-10. Here, $a_\mathrm{cen}$ denotes the inertial 
acceleration, $a_\mathrm{prs}$ the acceleration due to the thermal pressure 
gradient, and $a_\mathrm{lor}$ the acceleration by the magnetic pressure 
gradient.}
    \label{fig:streamline_xb10_forces}
\end{figure}

\begin{figure}[h]
    \centering
    \includegraphics[width=\linewidth]{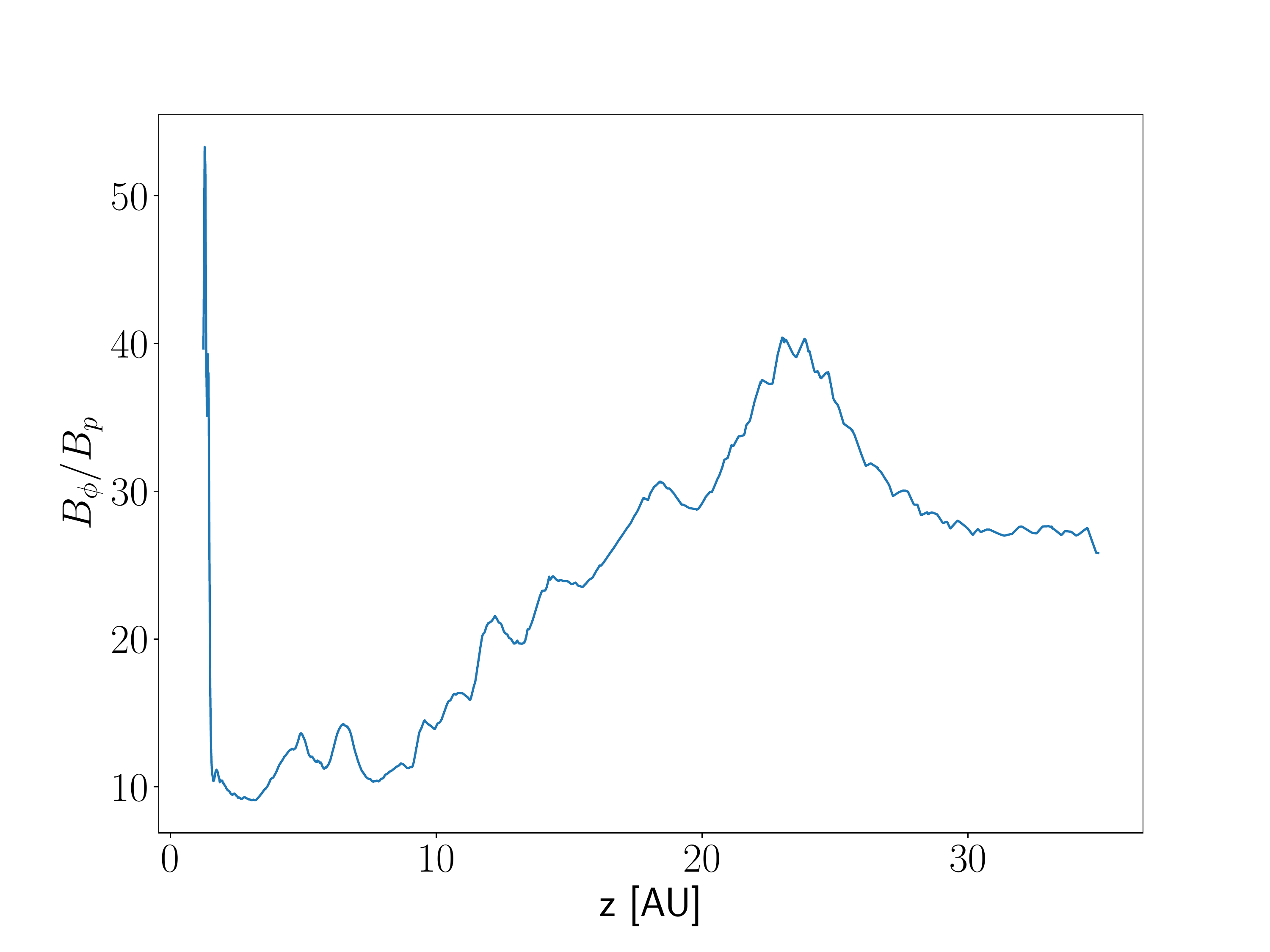}
    \caption[Poloidal velocities in X-b5]{Ratio of the toroidal and the poloidal 
magnetic field along a streamline in the time-averaged flow of X-b5. The 
toroidal field is clearly stronger by about one order of magnitude throughout 
the whole flow in the domain.}
    \label{fig:streamline_xb5_torfield}
\end{figure}

We focus on acceleration along the streamline. Fig. 
\ref{fig:streamline_xb5_forces} and Fig. \ref{fig:streamline_xb10_forces} 
provide further insight. In the context of the total acceleration $a_p$ , the 
contributing terms are (motivated by \cite{Bai2015b})
\begin{equation}
a_\mathrm{p} = - a_\mathrm{lor} - \frac{1}{\rho} \frac{\mathrm{d}P}{\mathrm{d}s} 
+ \left[ \frac{v_\phi^2}{r} \frac{\mathrm{d}r}{\mathrm{d}s} - \frac{\mathrm{d} 
\Phi}{\mathrm{d} s} \right]
,\end{equation}
where $\mathrm{d s}$ is an infinitesimally small line segment, tangential to the local 
streamline. 
The first term represents the acceleration by the Lorentz force 
$a_{\mathrm{lor}}$ in the diagrams, whereas the acceleration by the pressure 
force corresponds to $a_{\mathrm{prs}}$.
The third term can be considered as the excess of the centrifugal over the 
gravitational acceleration and is labeled $a_\mathrm{cen}$ in the plots. \\
The magnetic term $a_\mathrm{lor}$ can be expressed in the following way:
\begin{equation}
a_{\mathrm{lor}} = - \frac{1}{4 \pi \rho} \left( \frac{B_\phi}{r} \frac{\partial 
(r B_\phi)}{\partial r} + B_\phi  \frac{\partial B_\phi}{\partial z}  \right)
.\end{equation}
 The simplification made here is to neglect the possible magnetic tension forces 
contributed by the poloidal field. 
This approach is valid, as can be examined in Fig. 
\ref{fig:streamline_xb5_torfield}. Interestingly, the toroidal field is 
significantly stronger than the poloidal component for the whole streamline. 
Because a plasma beta of $10^5$ still corresponds to a relatively weak field, the 
shearing rotation strongly winds up the poloidal field. The approximation 
of only considering the magnetic pressure gradient originating from the 
azimuthal field is therefore reasonable.
Additionally, no forces due to the poloidal pressure gradient are considered 
here.\\
\begin{figure}[ht!]
    \centering
    \includegraphics[width=0.9\linewidth]{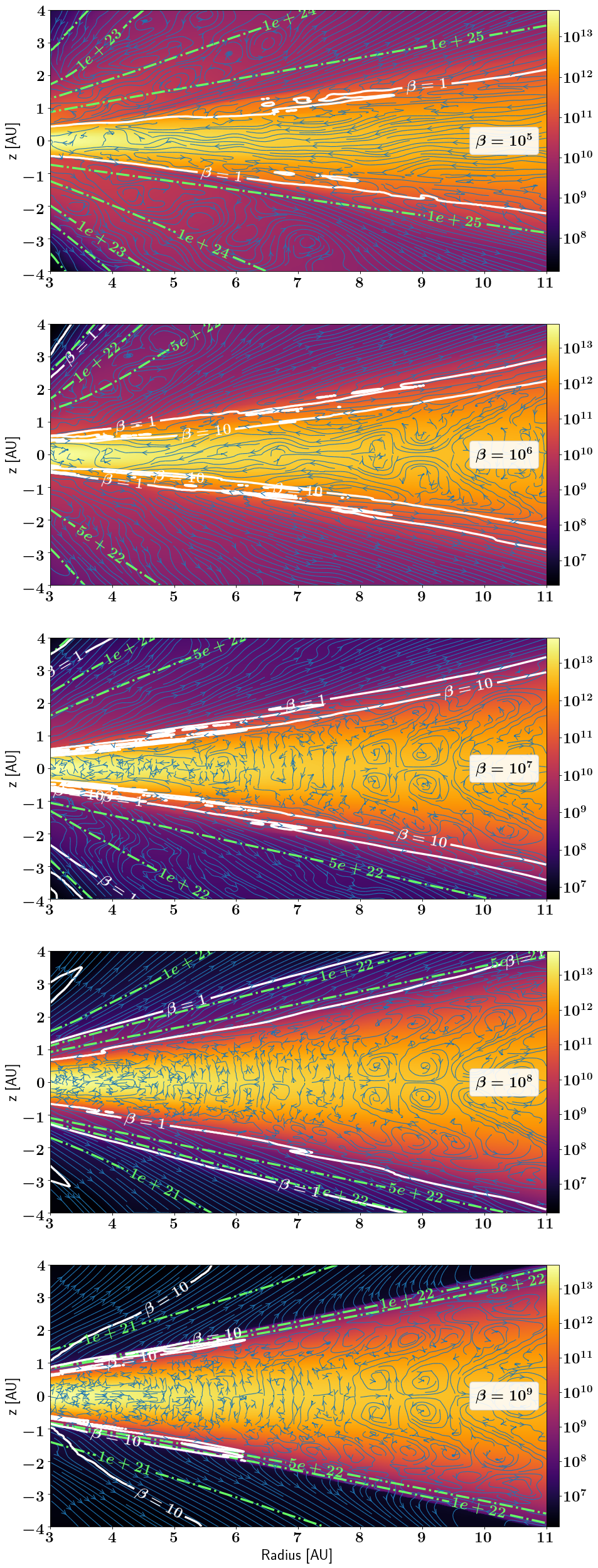}
    \caption{Time-averaged flows with varying initial 
values of $\beta$ in the mid-plane. The green dash-dotted line represents the radial 
column density in $\mathrm{cm}^{-2}$. The white lines mark contours of the local 
plasma beta. In all plots the blue arrowed lines trace velocity stream lines of 
the flow. The color map indicates the number density of the gas.}
    \label{fig:transition}
\end{figure}
\begin{figure}[ht]
    \centering
    \includegraphics[width=\linewidth]{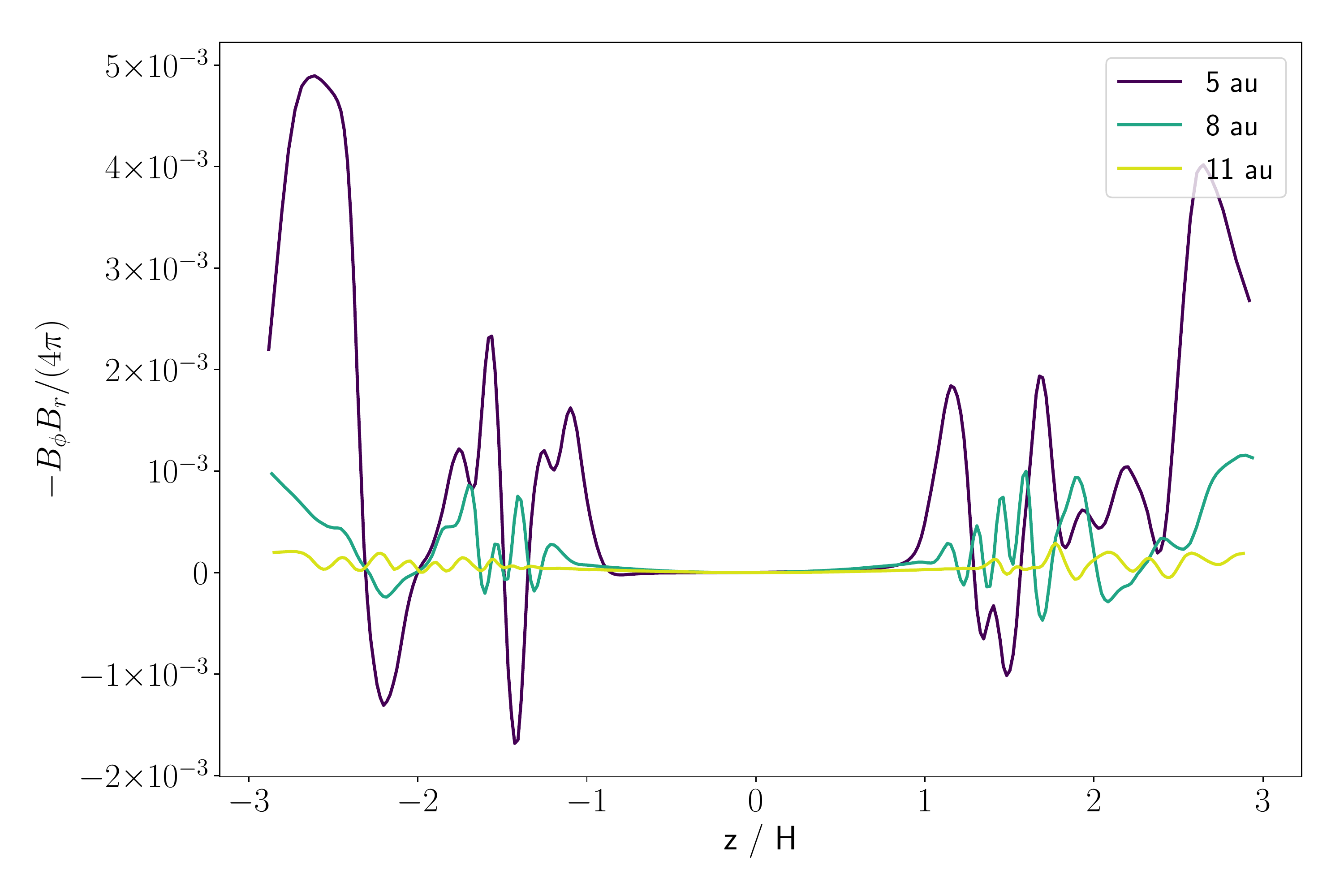}
    \caption{Tree lines represent the radial Maxwell stresses in cgs units. The stresses are mostly positive and  stronger toward inner radii and in the upper layers of the disk. In the mid-plane stresses are suppressed by ohmic diffusion.}
    \label{fig:stresses_radial}
\end{figure}
\begin{figure}[ht]
    \centering
    \includegraphics[width=\linewidth]{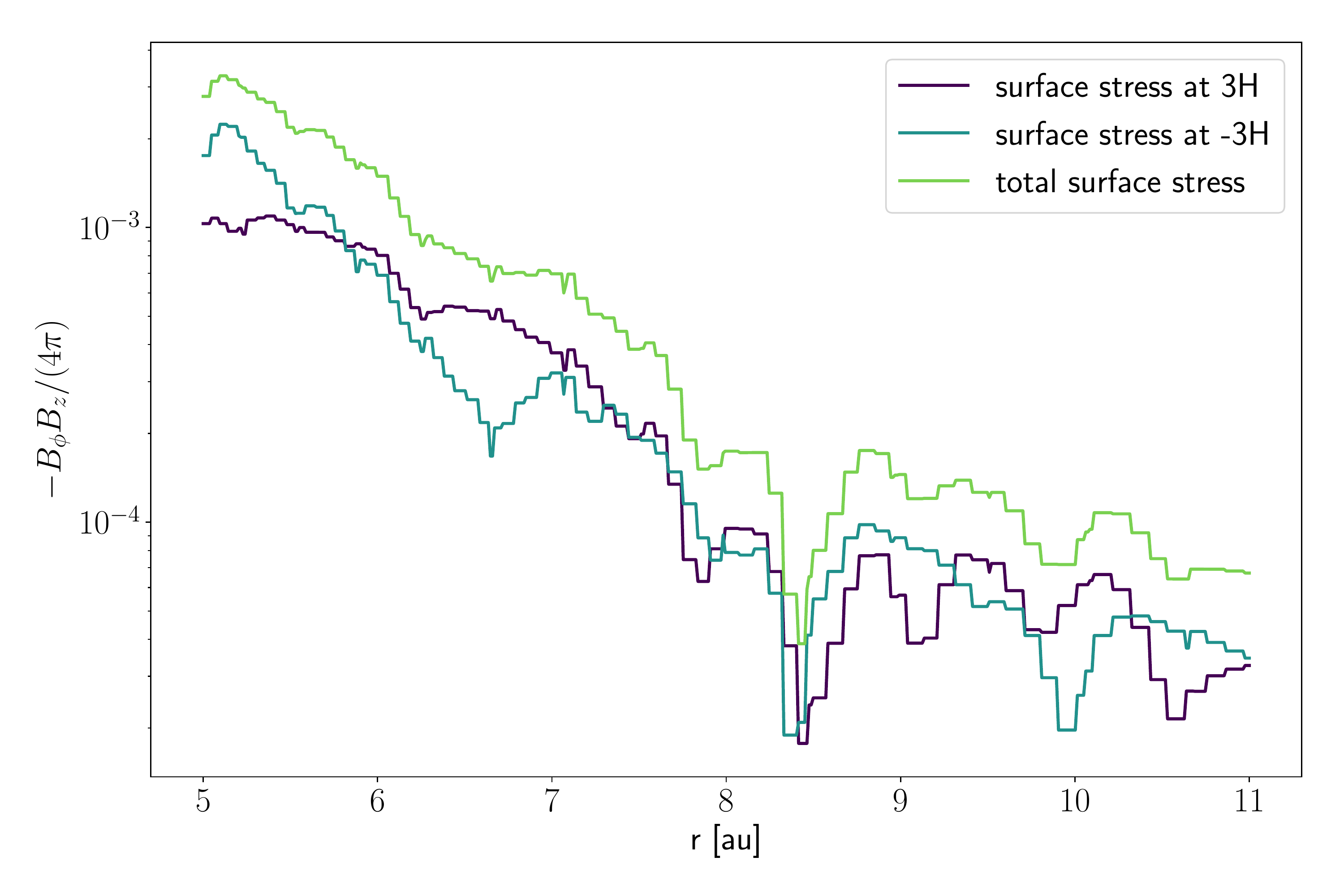}
    \caption{Magnetic surface stresses (cgs units) are evaluated at three pressure scale heights for a range of 5 au to 11 au. The lower and upper surface and their sum are represented by the three lines.}
    \label{fig:stresses_surface}
\end{figure}
\revised{For simulation X-b5, the magnetic acceleration effects are clearly visible along the specific streamline.} Additionally, $a_\mathrm{cen}$ is negative throughout the whole flow along the 
streamline as a result of the sub-Keplerian motion of the gas in the upper layers of the 
disk atmosphere where the wind-launching front is located. \revised{This leads to the 
conclusion that magnetocentrifugal acceleration is not present and that the 
magnetic and thermal pressure gradients are the driving factors for wind acceleration. }
The same conclusion was drawn in \cite{Bai2017b}. A reason for this phenomenon is 
the relatively weak poloidal field, which cannot sustain corotation to enforce 
magnetocentrifugal acceleration. 
Moreover, the Alfv\'en point, located closely to the wind-launching front, is 
consistent with this view.\\
The picture changes when the magnetic pressure is weaker than thermal pressure. As an example, accelerations along a streamline of the flow in run X-b10 with an initial $\beta = 10^{10}$ are shown 
in Fig. \ref{fig:streamline_xb10_forces}. The dominant acceleration mechanism is 
the thermal pressure gradient, and magnetic contribution is negligible 
throughout the whole streamline.
The result is consistent with the previous 
discussion and wind rates because simulation X-b10 is essentially equivalent to 
a photoevaporation run without magnetic fields. \\

\subsection{Analysis of the transition region} \label{sec:transition}
A further analysis of the transition between the magnetically and thermally
dominated wind regime is given in this section.  
Fig. \ref{fig:transition} shows radial column density contours as well as contours 
of the local plasma beta. 
Because the density of the magnetically launched wind decreases with increasing 
$\beta,$ the critical column density considered for the photoevaporation heating 
($5 \cdot 10^{22} \mathrm{cm}^{-2}$) moves closer to the wind-launching 
region in the upper disk layers. The launching region roughly coincides with the 
$\beta = 1$ surface when the wind is magnetically driven. Within a range of $10^7 
\leq \beta \leq 10^8$ , the critical column density lies below the upper $\beta=1$ 
surface. In the case of 
$\beta = 10^8$ , the magnetic field is too weak to eject material to regions far 
from the disk surface. The two surfaces of $\beta = 1$ are therefore located close 
to each other for each hemisphere. In regard to this transition occurring at the 
same magnetic field strengths as the saturation of the wind rates toward the 
photoevaporation rates, we can postulate the following:
When the $\beta=1$ surface is located below the critical photoevaporation 
column density, the wind is predominantly magnetically driven and drives higher 
mass-loss rates than the thermally driven wind. 
When the critical photoevaporation column density lies below the $\beta = 
1$ surface, however, the wind can be considered to be mainly thermally driven, 
with a wind rate equal to the photoevaporation rate, and not being dependent on the 
magnetic field strength in this regime. \\
It might be argued that the decreasing wind rate in the magnetic regime could be 
caused by a lack of sufficient numerical resolution with respect to the most 
unstable MRI mode. However, the wind launching should not rely on MRI because
it is launched by the thermal and magnetic pressure gradient (see sec. 
\ref{sec:flow_analysis}). \\
In the following we justify this argument. The wavelength of the fastest 
growing MRI mode $\lambda_{\mathrm{max}}$ is given by $\lambda_{\mathrm{max}} = 
\frac{4}{\sqrt{15}} \, v_\mathrm{Az} / \Omega_\mathrm{K}$ \citep{Balbus1991} with the vertical 
Alfv\'en velocity $v_\mathrm{Az}$. We define the quality factor $q_\mathrm{MRI} 
= \lambda_\mathrm{max} / \Delta z$. For values of $\beta$ of $(10^5; 10^7; 10^{10}),$ 
this ratio becomes $(4.95, 0.495, 0.0156)$. The corresponding local values of 
the primitive variables are taken at a height of $3 H$, which is located 
approximately at the wind-launching front. The numerical resolution is 
therefore clearly not sufficient to resolve effects of the MRI throughout the whole 
parameter range. \\
In order to justify the independence of the wind-launching 
mechanism of these numerical caveats, we performed a simulation with the same 
initial parameters as run X-b6, but with a lower resolution of 200x200 in 
radius and polar angle (X-b6-mri). The radial grid was again logarithmically scaled, but we 
chose a uniformly spaced grid in polar direction. With this setup, the quality 
factor becomes $q_\mathrm{MRI} \approx 0.3$. \\
Applying the same averaging and 
wind rate calculation as for the previous runs, we obtained a mass-loss rate of 
$(6.24 \pm 2.56) \cdot 10^{-8} M_\odot \mathrm{yr}^{-1}$. Given the low 
resolution, the result agrees well with the mass-loss rate of X-b6. We can thus 
conclude that the numerical resolution of MRI modes cannot account for 
the transition between magnetically driven disk winds and photoevaporation in 
our simulations.

\subsection{Accretion rate and lever arm} \label{sec:accretion_rate}
In Table \ref{tab:windrates} we list the accretion rates measured for all simulation runs, including magnetic fields. The rates computed from time-averaged flows with a radial mean range from 5 au to 10 au. Clearly, the accretion rates are significantly lower than the corresponding wind rates. Especially for $\beta \geq 10^7$ the ratio increases to more than one order of magnitude. In contrast to $\beta = 10^5$ and $\beta = 10^6$ , the streamlines in Fig. \ref{fig:transition} show no coherent accretion flow at values of $\beta$ beyond the transition toward a thermally driven wind. For stronger fields the accretion flow dissolves in the region within $\approx 3 - 4 \, \mathrm{au,}$ where a more complex and stronger toroidal magnetic field is present (see Fig. \ref{fig:field_toroidal}).\\
Accretion rates lower than the wind rates were also observed by \cite{Gressel2015b}. Overall, 
the authors measured higher wind rates than accretion rates. A similar picture emerged in \cite{Bai2017b}. In his model the accretion rate was about $2 \cdot 10^{-8} M_\odot \mathrm{yr}^{-1}$. The difference between the results of these two works and ours could be attributed to the alternative approach in the chemical model and the elevated ionization rate, as well as to the restriction to ohmic diffusion.\\
We verified that both radial and vertical Maxwell stresses were present in the simulations. As an example, the radial Maxwell stresses are plotted in Fig. \ref{fig:stresses_radial} for different positions along the radial direction. The stresses are most prominent in the upper surface layers beyond one scale height. Closer to the mid-plane, ohmic diffusion inhibits these stresses. In Fig. \ref{fig:stresses_surface} the vertical Maxwell stresses at three scale heights are displayed for the upper and lower surface. Because the magnetic field structure is rather complex, a calculation of the accretion rate by both radial and vertical stresses, taken at the positions here, would be unreliable. However, the stresses indicate that the magnetic field indeed transports momentum in the strong-field simulations.
For simulation X-b5, we measured the magnetic lever arm. Streamlines were traced starting from three scale heights until the poloidal gas velocity became super-Alfv\'enic. The procedure was carried out over a region of 5.5 au to 12 au, resulting in a lever arm of $\lambda = 1.06 \pm 0.03$. \cite{Bai2017b} reported a value of 1.15 for the same $\beta = 10^5$. Fig. \ref{fig:streamline_xb5} and Fig. \ref{fig:streamline_xb5_zoom} confirm that the wind basically starts super-Alfv\'enic, leading to the small lever arm. \\
\section{Discussion} \label{sec:discussion}
The imposed parameter space of $\beta$ is sufficient to display the two regimes 
of photoevaporation and magnetically driven disk winds, indicated by the mass 
that is carried away through the wind and by the accelerations in Fig. 
\ref{fig:streamline_xb5_forces} and Fig. \ref{fig:streamline_xb10_forces}.
For a lower plasma beta, no completely stationary flow configuration is reached, 
which is represented by the evolution of the mass-loss rates in Fig. 
\ref{fig:windfluxes}. \\
The irregular surface structure prevents a clear definition of a wind-launching 
front. As a proxy, the $\beta = 1$ surface and the region at $3 H$ were used.
Ambipolar diffusion was observed to lower the wind mass-loss rates and to 
dampen the upper layers of the atmosphere \citep{Gressel2015b}.
The transition region between magnetically driven winds and photoevaporation 
would then shift toward stronger magnetic field strengths.
 In our simulations we did not include this effect because we aimed to extend the 
current photoevaporation model by \cite{Picogna2019} step by step. We reserve 
the inclusion of ambipolar diffusion and the Hall effect to future studies.\\
To compare the results with previous works, our mass-loss rates are on the same 
order of magnitude as those described by \cite{Bethune2017}. However, the rates measured 
in our work are higher by a factor 3 - 5. This difference most likely 
originates from the exclusion of ambipolar diffusion here. Because B\'eth\'ethune and collaborators did not 
include grains in their chemical model, the ionization fraction within the disk 
is significantly lower in our case. A similar complex field structure emerged in 
the simulation of \cite{Gressel2015b} without ambipolar diffusion. 
Their initial ohmic Elsasser number profile is comparable to that in our work. \\
The flow also resembles the results of simulations with $\beta = 5000$ in 
\cite{Sheikhnezami2012}, where for this value of the plasma parameter, 
magnetocentrifugal effects play a minor role. The authors approached the 
Blandford-Payne regime for smaller $\beta$ down to 10. Thus, the minimum plasma 
parameter of $10^5$ applied in our simulations lies far from the transition to 
magnetocentrifugal winds. The observed outflows in our work do not provide a 
significantly large lever arm ($\lambda \approx 1.06$). \\
Radial and vertical stresses are present in our models, however, and drive relatively weak accretion flows compared to the wind rates in the strong-field limit. The accretion rates for $\beta \geq 10^8$ saturate to a low level of $\approx 2\cdot 10^{-10} M_\odot \mathrm{yr}^{-1}$. These rates are most likely of numerical origin because no explicit viscosity is applied in the simulations. Numeric diffusion might lead to these nonzero net fluxes. \\
 \cite{Bai2017b} used a similar plasma beta range and identified the magnetic 
pressure gradient as the main wind-launching mechanism.
They reported that by applying an increasingly strong magnetic field on a purely 
thermal disk wind, a transition from magnetic pressure driven to 
magneto-centrifugally driven winds is to be expected. 
Very recently, \cite{Wang2019} introduced a consistent thermochemical model to 
the framework of \cite{Bai2017b}. 
In contrast to the simulations performed here, no wide plasma parameter sweep was conducted. The measured wind rates are about one order of magnitude lower 
than in model X-b5. \\
In the 2D framework, MRI activity is 
not realistically represented, especially the development of the toroidal 
magnetic field. A proper study in three dimensions is required 
to follow the evolution more accurately.
However, the inclusion of ambipolar diffusion might suppress MRI in 
the upper layers (or at least provide a more steady wind), as also observed by the previously mentioned authors. \\
We based our focus on EUV and X-ray radiation as the driving component for 
photoevaporation on the findings of \cite{Ercolano2009}. They argued that heating 
by X-ray photons is dominant and leads to higher mass-loss rates than EUV 
heating. \\

\section{Conclusion} \label{sec:conclusion}
We presented simulations that systematically studied the transition region between 
photoevaporation and magnetically driven disk winds involving a fully global 
2.5D axisymmetric and non-ideal magnetohydrodynamic framework. Our model includes
heating by EUV and X-ray radiation and ionization by direct and scattered X-rays as 
well as cosmic rays.
The main results can be summarized as follows:
\begin{enumerate}
\item \revised{A magnetothermal wind without magnetocentrifugal acceleration emerges} 
for a plasma beta of $\beta = 10^5$ to $10^7$, leading to wind rates of about $10^{-8} - 10^{-7} M_{\odot} \mathrm{yr}^{-1}$. 
\item The transition region is identified to be in the range of $\beta \geq 10^7$, 
where the wind rates deviate from a power-law dependence on the initial plasma 
beta and saturate toward the pure photoevaporation rate.
The transition occurs when the critical photoevaporation column density approaches the $\beta = 1$ surface. This surface roughly corresponds to the wind-launching surface in the magnetically driven wind scenario. For $\beta = 10^5 - 10^6$ the wind is optically thick with respect to the ionization heating. When $\beta > 10^7$ , the critical photoevaporation column density surface lies below the region of $\beta = 1$ and the wind becomes dominantly thermally driven.
\item  The photoevaporation flow is sensitive to small perturbations caused by 
the magnetic field in the upper disk layers. Fluctuations in these regions 
temporarily alter the radial column densities and prevent the surface layers in 
the outer region from being heated by the EUV and X-ray radiation.
The photoevaporation rate without magnetic fields is $(9.66 \pm 0.44) \cdot 
10^{-8} M_{\odot} \mathrm{yr}^{-1}$.
\end{enumerate}
Because magnetic field strengths for observed stellar systems are still uncertain \citep{Fang2018, Vlemmings2019}, it is still difficult to evaluate
the importance of magnetic versus photoevaporative winds. Our results show that the transition between these two wind-launching mechanisms strongly depends on the magnetization of the disk. Our work serves as
motivation for better measurements of B fields, however. Alternatively, when more wind fluxes will
have been measured, the underlying values of $\beta$
may be reverse-engineered based on our simulations.

\begin{acknowledgements}
P. R. wants to thank Giovanni Picogna and Barbara Ercolano for helpful 
discussions regarding the ionization and heating model. Additionally, P. R. 
thanks Giovanni Picogna for providing the X-ray photoionization temperatures.
P. R. acknowledges the support of the DFG Research Unit `Transition Disks' (FOR 
2634/1, DU 414/23-1).
The simulations for the project were performed on the ISAAC cluster owned by the 
MPIA and the HYDRA and DRACO clusters of the Max-Planck-Society, both hosted at 
the Max-Planck Computing and Data Facility in Garching (Germany).

\end{acknowledgements}

\bibliographystyle{aa}
\bibliography{library}

\end{document}